\documentclass[utf8]{frontiersSCNS} 

\usepackage{url,lineno,microtype,subcaption}
\usepackage[onehalfspacing]{setspace}
\usepackage{amsmath,amssymb}
\usepackage{mathrsfs}

\newcommand{\eg}{{{e.g.}}}

\newcommand{\PD}[2]{\partial_{#2} #1}
\newcommand{\vet}[1]{ {\bf{#1}}}

\newcommand{\ffrac}[2]{\ensuremath{\frac{\displaystyle #1}{\displaystyle #2}}}

\renewcommand{\vec}[1]{\boldsymbol{#1}}

\newcommand{\rmi}{{{\mathrm{i}}}}

\newcommand{\mcR}{{{\mathcal{R}}}}
\newcommand{\wmcR}{{\mathcal{\widehat{R}}}}
\newcommand{\mcL}{{\mathcal{L}}}
\newcommand{\bA}{\boldsymbol{A}}

\newcommand{\bO}{\vet O}
\newcommand{\bI}{\vet I}
\newcommand{\bvO}{\vet{\widehat{O}}}
\newcommand{\bR}{{\vet{R}}}
\newcommand{\br}{\vet r}
\newcommand{\bz}{\vet z}

\newcommand{\ie}{{{i.e.}}}
\newcommand{\be}{{\mbox{\boldmath${e}$} }}
\newcommand{\bxi}{{\mbox{\boldmath${\xi}$} }}
\newcommand{\bfeta}{{\mbox{\boldmath${\eta}$} }}

\newcommand{\Ylm}{{Y_\ell^m}}

\newcommand{\Hz}{{\mbox{$\mathrm{Hz}$}}}
\newcommand{\nHz}{{\mbox{$\mathrm{nHz}$}}}
\newcommand{\df}{\mathrm{d}}

\newcommand{\rl}{r^{(\ell)}}

\newcommand{\mcN}{{{\mathcal{N}}}}

\newcommand{\OO}{{{\mathcal{O}}}}
\newcommand{\mc}[1]{{{\mathscr{#1}}}}
\newcommand{\RM}[1]{{{\mathrm{#1}}}}
\newcommand{\inner}[2]{\left \langle {#1},{#2}\right \rangle}

\renewcommand{\Re}{\mathrm{Re}}
\newcommand{\beq}{\begin{equation}}
\newcommand{\eeq}{\end{equation}}
\newcommand{\beqal}{\begin{align}}
\newcommand{\eeqal}{\end{align}}

\def\keyFont{\fontsize{8}{11}\helveticabold }
\def\firstAuthorLast{Papini and Gizon}
\def\Authors{Emanuele Papini\,$^{1}$ and Laurent Gizon\,$^{1,2,*}$}
\def\Address{$^{1}$Max-Planck-Institut f\"ur Sonnensystemforschung, G\"ottingen, Germany \\
$^{2}$Institut f\"ur Astrophysik, Georg-August-Universit\"at G\"ottingen, G\"ottingen, Germany  }
\def\corrAuthor{Max-Planck-Institut f\"ur Sonnensystemforschung, Justus-von-Liebig-Weg 3, 37077 G\"ottingen, Germany}
\def\corrEmail{gizon@mps.mpg.de}

\begin{document}
\onecolumn
\firstpage{1}

\title[Asteroseismic Signature of a Large Active Region]{Asteroseismic Signature of a Large~Active~Region} 

\author[\firstAuthorLast ]{\Authors} 
\address{} 
\correspondance{} 

\extraAuth{}

\maketitle

\begin{abstract}

\section{}
Axisymmetric magnetic activity on the Sun and sun-like stars increases the frequencies of the modes of acoustic oscillation. However, it is unclear how a corotating patch of activity  affects the oscillations, since such a perturbation is unsteady in the frame of the observer. In this paper we qualitatively describe the asteroseismic signature of a large active region in the power spectrum of the dipole ($\ell=1$) and quadrupole ($\ell=2$)  p modes.
First we calculate the frequencies and the relative amplitudes of the azimuthal modes of oscillation in a frame that corotates with the active region, using first-order perturbation theory. For the sake of simplicity, the influence of the active region is approximated by a near-surface increase in sound speed.  In the corotating frame the perturbations due to (differential) rotation and the active region completely lift the ($2\ell+1$)-fold azimuthal  degeneracy of the frequency spectrum of modes with harmonic degree $\ell$. 
Then we transform to an inertial frame to obtain the observed power spectrum. In the frame of the observer, the unsteady nature of the perturbation leads to the appearance of $(2\ell+1)^2$ peaks in the power spectrum of a multiplet. These peaks blend into each other to form asymmetric line profiles. In the limit of a small active region (angular diameter less than $30^\circ$), we approximate the power spectrum of a multiplet in terms of $2\times(2\ell+1)$ peaks, whose amplitudes and frequencies depend on the latitude of the active region and the inclination angle of the star's rotation axis.
In order to check the results and to explore the nonlinear regime, we perform numerical simulations using the 3D time-domain pseudo-spectral linear pulsation code GLASS. 
For small sound-speed perturbations, we find a good agreement between the simulations and linear theory. Larger perturbation amplitudes will induce mode mixing and lead to additional complex changes in the predicted power spectrum.
{  However linear perturbation theory provides useful guidance to search for the observational signature of large individual active regions in stellar oscillation power spectra.}

\tiny
 \keyFont{ \section{Keywords:} asteroseismology, stars: oscillations, stars: activity, stars: rotation, starspots} 
\end{abstract}

\section{Introduction}

Surface magnetic activity shifts the frequencies of the global modes of acoustic oscillation during solar and stellar activity cycles  \citep[e.g.,][]{Palle1989,Garcia2010,Santos2016,Kiefer2017}. Asteroseismology can in turn inform us about the strength and the latitude distribution of a band of magnetic activity on the stellar surface  \citep{Gizon2002,Chaplin2007a,Gizon2016}. However, the signature of a single large active region in stellar p-mode oscillation power spectra has not been discussed so far in detail.  A complication inherent to this problem comes from the fact that the perturbation associated with a rotating active region is not steady in the observer's frame. 
Yet, this problem is relevant, since large starspots are detected in photometric variability \citep[\eg,][]{Mosser2009a} and using Doppler or Zeeman-Doppler imaging  \citep[\eg,][]{Strassmeier2009}.

We build on preliminary work by \citet{gizon1995, gizon1998} who considered the effect of a single sunspot in corotation on the low-degree modes of solar oscillation. 
The problem of unsteady perturbations has been considered in the past in different contexts. The interaction with acoustic modes with an inclined magnetic field with respect to the stellar rotation axis was first studied by \citet{Kurtz1982} to explain the oscillation power spectra of roAp stars \citep[see also][]{Kurtz2008}.
In the oblique pulsator model the effect of the magnetic field dominates over rotation, and the pulsation axis is aligned with the magnetic axis of the star. Only modes of oscillations symmetric with respect to the magnetic field axis are excited.
The oblique pulsator model was extended by \citet{Dziembowski1985} to include the first-order effects of the Coriolis and Lorentz forces, and then  by \citet{Shibahashi1993}  to account for the distortion in the eigenfunctions.
In parallel \citet{Dziembowski1984} and \citet{Gough1984}  discussed the combined influence of rotation and of an inclined magnetic field in corotation  on multiplets of solar acoustic oscillations. They explicitly mentioned that each multiplet consist of $(2\ell+1)^2$ components in the power spectrum \citep[this was already hinted at by][]{Dicke1982}.

We focus on stars with a level of activity higher than the Sun, which may have active regions with larger surface coverage, and therefore better chances for detection. Following the same approach of  \citet{goode1992}, we investigate the linear changes  induced in a $n\ell$-multiplet by an unsteady perturbation, that mimics an active region (AR) rotating with the star. In particular we study the power spectra of the dipole and quadrupole multiplets. For the active region we consider a simple two-parameter model, where near-surface sound-speed perturbations have a given amplitude and surface coverage.

As a complement to the linear analysis we also explore the nonlinear regime of the active-region perturbation by means of 3D numerical simulations, by studying the combined effect of rotation and mode mixing on the observed power spectra using the wave propagation code GLASS \citep[see][]{Hanasoge2007,Papini2015}. 
{ We note that the non-linear regime was studied in the context of strong magnetic fields in roAp stars by, e.g., \citet{Cunha2000},  \citet{Bigot2002}, \citet{Saio2004}, and  \citet{Cunha2006}.}

\section{Methods}

\subsection{Signature of an active region in the oscillation power spectrum: linear theory}
\label{cha:rotation:sec:lineartheory}

\subsubsection{Linear problem in the corotating frame}

The normal modes of oscillation of a spherically symmetric non-rotating star are identified by three integer numbers: the radial order $n$, the angular degree $\ell$, and the azimuthal order $m$, with $|m|\le \ell$.
In the absence of attenuation the degenerate mode frequencies, $\omega_{n\ell}^{(0)}$, are real and the displacement eigenvectors $\bxi^{(0)}_{n\ell m}(\br) \exp(-\rmi \omega_{n\ell}^{(0)} t)$ solve
the linearized equation of motion
\begin{gather}
\label{eq:H0xi0}
 \mcL^{(0)} [ \bxi_{n\ell m}^{(0)} ]  = \omega_{n\ell}^{(0)2} \bxi_{n\ell m}^{(0)} ,
\end{gather}
where $\mcL^{(0)}$ is a linear spatial differential operator \citep[see, \eg, ][]{Unno1979}. 
{ Hereafter superscripts ''(0)'' denote quantities associated with the non-rotating stellar model.}
In spherical polar coordinates $\br=(r,\theta,\phi)$, the displacement eigenfunctions can be written as
\begin{gather}
\label{eq:xi_0}
 \bxi_{n\ell m}^{(0)}(\br)= 
      \left [\xi_{r,n\ell}(r) \be_r +
      \xi_{h,n\ell}(r) \left ( \be_\theta \PD{ }{\theta} + 
             \frac{1}{\sin\theta}\be_\phi\PD{ }{\phi} \right )\right] Y_l^m(\theta,\phi) ,
\end{gather}
where $Y_l^m(\theta,\phi)$ are spherical harmonics and $\xi_{r,n\ell}$ and $\xi_{h,n\ell}$ are functions of radius only that can be calculated numerically for a given stellar model \citep[see, e.g.,][]{Aerts2010}.

We now consider two perturbations, the first perturbation due to rotation \citep[e.g.,][]{Hansen1977} and the second arising from the presence of an active region that rotates with the star \citep[see, e.g.,][for the effect of a sunspot on high-degree modes]{Schunker2013}.   The latter perturbation is unsteady in any inertial frame of reference. 
Here we aim to study how these two effects may affect the fine structure of the modes within a fixed multiplet $(n,\ell)$.
The two perturbations taken together completely remove the $(2\ell+1)$-fold degeneracy in $m$.
\begin{figure}
 \includegraphics[width=\columnwidth]{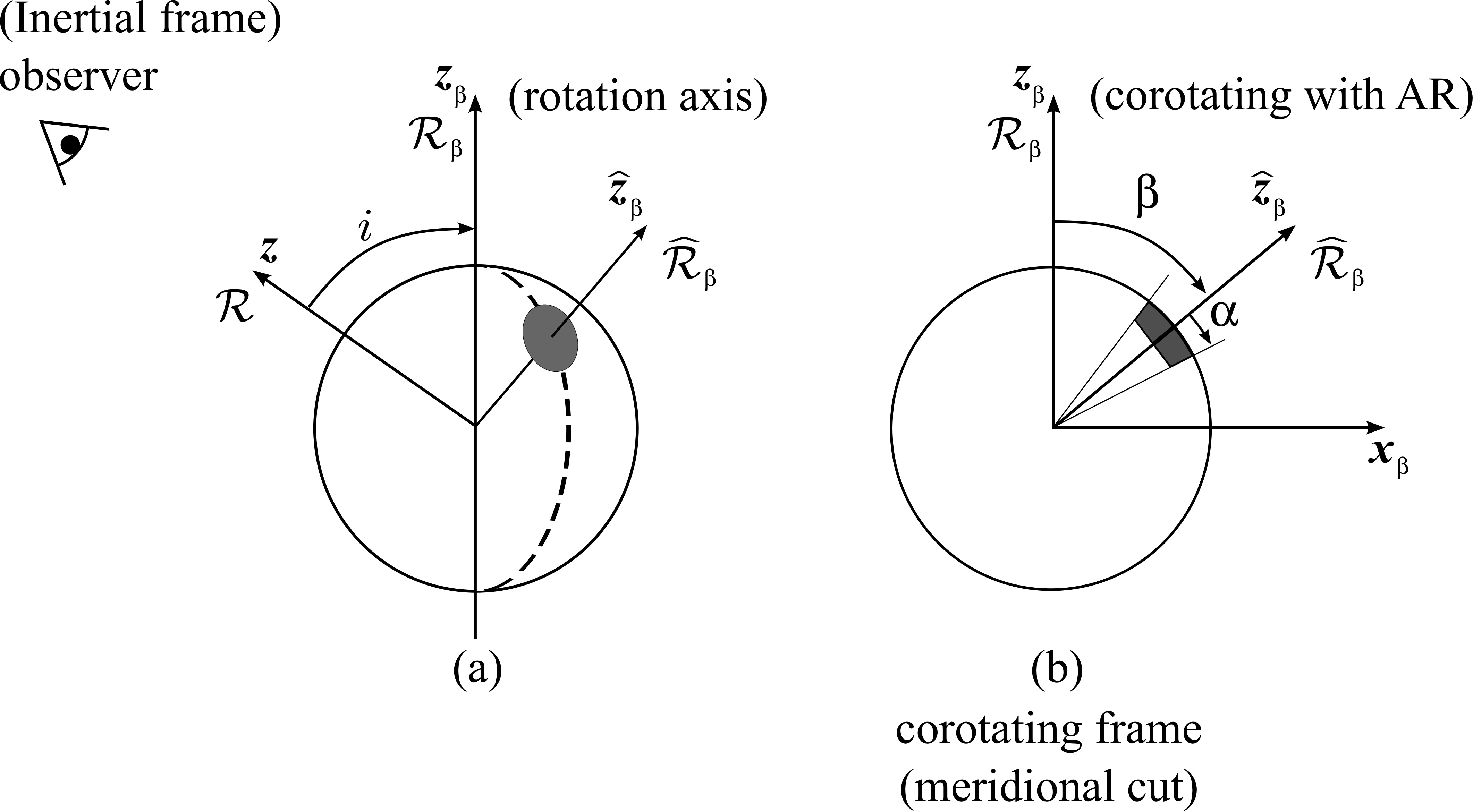}
 \caption[reference frames]{Reference frames and angles of the problem. Arrows show the polar axes of the coordinate systems $\mcR$, $\mcR_\beta$, and $\widehat\mcR_\beta$. The frame $\mcR$ is the inertial frame of the observer.
 The rotation axis of the star is inclined by an angle $i$ with respect to the line of sight. 
 Both frames $\mcR_\beta$ and $\widehat\mcR_\beta$ corotate with the active region (shaded area) at a constant angular velocity $\Omega_\beta$. The polar axis of $\mcR_\beta$ is aligned with the rotation axis of the star. In $\mcR_\beta$ the active region has a colatitude $\beta$ and in $\widehat\mcR_\beta$ it is at the pole. 
 }
 \label{fig:referenceframes}
\end{figure}

Provided that there is only one active region, it is much more convenient to first tackle the problem in a reference frame that is corotating with the active region, where both perturbations are steady \citep[e.g.,][]{Dziembowski1984,goode1992}. 
 In Fig. \ref{fig:referenceframes} we define three frames of reference, $\mcR$, $\mcR_\beta$, and $\widehat{\mcR}_\beta$, all three with the same origin at the center of the star.
Frame $\mcR$ is an inertial frame of reference, with polar axis $\bz$ pointing towards the observer. 
We denote by $\beta$ the colatitude of the active region in frame $\mcR$. The other two frames are both corotating with the active region at the angular velocity $\Omega_\beta$ about the rotation axis of the star. 
The polar axis of $\mcR_\beta$ is directed along the stellar rotation axis and is inclined by an angle $i$ with respect to $\bz$. The polar axis of the frame $\widehat\mcR_\beta$ is inclined by the angle $\beta$ with respect to the rotation axis. In $\wmcR_\beta$ the center of the active region is at the north pole. 
Provided that the starspot has bno proper motion, the angular velocity $\Omega_\beta$ is equal to the surface rotational angular velocity of the star at  colatitude $\beta$.
We call $\br=(r,\theta,\phi)$, $\br_\beta=(r,\theta_\beta,\phi_\beta)$, and $\widehat\br_\beta=(r,\widehat\theta_\beta,\widehat\phi_\beta)$ the spherical-polar coordinates associated with $\mcR$, $\mcR_\beta$, and $\wmcR_\beta$ respectively.

We consider the effects of rotation and the active region on the acoustic oscillations as small perturbations. In the frame $\mcR_\beta$ each mode is identified by the index $M$, $-\ell \leq M \leq \ell$, and we expand the displacement eigenvectors and eigenfunctions as 
\begin{equation}
\label{eq:xi_pert}
\bxi_{n\ell M} (\br_\beta) = \sum_{m=-\ell}^{\ell} A^{M}_{m}  \bxi^{(0)}_{n\ell m} (\br_\beta) + \delta\bxi_{n\ell M} (\br_\beta) +\cdots 
\end{equation}
and 
\begin{equation}
\label{eq:freq_pert}
\omega_{n\ell M} = \omega^{(0)}_{n\ell} + \delta \omega_{n\ell M} + \cdots ,
\end{equation}
where $\delta\bxi_{n\ell M}$ is orthogonal to each unperturbed eigenvector $\bxi^{(0)}_{n\ell m}$ with the same $\ell$ and $n$ \citep[\eg][]{1990gough}, 
and { the coefficients $A^{M}_{m}$ are (real) amplitudes}.
We write the wave operator as
\begin{equation}
\mcL = \mcL^{(0)} + \delta \mcL  + \cdots, 
\end{equation}
with
\begin{equation}
\delta \mcL  =   \mcL_\Omega +  \mcL_{\rm AR},
\end{equation}
where $\mcL_{\Omega}$ accounts for  the effects of rotation and $\mcL_{\rm AR}$  for the effects of the active region.
To first order, the linearized equation of motion reduces to 
\begin{equation}
  \sum_{m=-\ell}^\ell A_{m}^M \left (\mcL_\Omega+\mcL_\RM{AR} \right )[\bxi^{(0)}_{n\ell m}]  + \mcL^{(0)} [ \delta\bxi_{n\ell M} ] = 
 2 \omega^{(0)}_{n\ell}  \delta \omega_{n\ell M}  \sum_{m=-\ell}^\ell  A^{M}_{m}  \bxi^{(0)}_{n\ell m}  +     \omega_{n\ell}^{(0)2} \delta\bxi_{n\ell M}.
\label{eq:HMxiM}
\end{equation}
We define the inner product between two vectors $\bxi(\br_\beta)$ and $\bfeta(\br_\beta)$ on the Hilbert space of displacement vectors as
\begin{gather}
 \inner{\bxi}{\bfeta} =\int_V \bxi^* \cdot \bfeta\,\rho\df V  ,
\end{gather}
where $^*$ denotes the complex conjugate and $V$ is the stellar volume.
The unpertubed eigenmodes are normalized such that
$$\inner{\bxi^{(0)}_{n\ell m'}}{\bxi^{(0)}_{n\ell m}}=\delta_{m'm}.$$
We take the inner product of Eq. (\ref{eq:HMxiM}) with $\bxi^{(0)}_{n\ell m'}$ to obtain
\begin{equation}
  \sum_{m=-\ell}^\ell A_{m}^M \inner{\bxi^{(0)}_{n\ell m'}}{ (\mcL_\Omega+\mcL_{AR})[\bxi^{(0)}_{n\ell m} ] }  
+ \inner{ \bxi^{(0)}_{n\ell m'} }{  \mcL^{(0)} [ \delta\bxi_{n\ell M} ] }  
=2 \omega^{(0)}_{n\ell}  \delta \omega_{n\ell M}   A_{m'}^M .   
\label{eq:inner1}
\end{equation}
Because $\mcL^{(0)}$ is Hermitian symmetric and $\langle  \bxi^{(0)}_{n\ell m'} ,\delta\bxi_{n\ell M} \rangle =0$, the second term on the left-hand side of the above equation vanishes
\begin{align*}
\inner { \bxi^{(0)}_{n\ell m'} } 
	   {\mcL^{(0)} [ \delta\bxi_{n\ell M} ] } 
        = \inner {\mcL^{(0)} [\bxi^{(0)}_{n\ell m'}] } 
       {   \delta\bxi_{n\ell M}  } =0 .
\end{align*}
Introducing the perturbation matrix elements
\begin{align}
O_{m' m} = O^{\Omega}_{m' m} + O^{\rm AR}_{m' m},
\end{align}
where
\begin{align}
& O^{\Omega}_{m' m} =   \frac{1}{2\omega^{(0)}_{n\ell}} 
\inner {\bxi^{(0)}_{n\ell m'} }{ \mcL_\Omega [ \bxi^{(0)}_{n\ell m} ]} \\
& O^{\rm AR}_{m' m}  =  \frac{1}{2\omega^{(0)}_{n\ell}} 
\inner {\bxi^{(0)}_{n\ell m'} }{ \mcL_{\rm AR} [ \bxi^{(0)}_{n\ell m} ] } ,
\end{align}
Equation (\ref{eq:inner1}) becomes
\begin{align}
    \sum_{m=-\ell}^\ell O_{m'm}  A^{M}_{m}      = 
   \delta \omega_{M}   A^M_{m'}.   
\end{align}
To simplify the notation we dropped the indices $n\ell$ on $\delta\omega_M$.
In matrix form,
\begin{align}
\label{eq:eigen_problem}
    \bO \vec{A}^{M} = 
   \delta \omega_{M}  \;  \vec{A}^{M},
\end{align}
where
$
\vec{A}^M= [A_{-\ell}^M \; A_{-\ell+1}^M \;\dots \; A_{\ell}^M]^T
$
is the vector of amplitudes. 
To find $\vec{A}^M$ and $\delta\omega_M $  we have to solve the above eigenvalue problem, Eq. (\ref{eq:eigen_problem}).

The rotation perturbation matrix $\bO^{\Omega}$ is diagonal in the frame $\mcR_\beta$.
The active region perturbation matrix $\bO^{\mathrm{AR}}$ is not diagonal in $\mcR_\beta$, 
 but it can be obtained in  terms of 
 the diagonal perturbation matrix ${\bvO}^{\mathrm{AR}}$ expressed in the frame $\wmcR_\beta$, 
\begin{gather}
\label{eq:OtildeAR_toOAR}
\bO^\mathrm{AR}=\bR^{(\ell)}{\bvO}^\mathrm{AR} 
({\bR^{(\ell)}})^{-1}. 
\end{gather}
where the matrix $\bR^{(\ell)}$ performs 
a clockwise rotation of $\beta$ about the $y$ axis that transforms the frame $\wmcR_\beta$ into the frame $\mcR_\beta$.
More explicitly, the elements of the rotation matrix are given by
\begin{gather}
\label{eq:rotylmabg}
R_{m\,m'}^{(\ell)}= r^{(\ell)}_{m\,m'}(-\beta) = r^{(\ell)}_{m'm}(\beta),
\end{gather}
where  $r^{(\ell)}_{m\,m'}(\beta)$ is given by \citet{1959messiah}.

\label{sec:corot}

\subsubsection{Frequency splittings due to rotation}

In the corotating frame $\mcR_\beta$ the rotation perturbation matrix is diagonal:
\begin{align}
 O^\Omega_{m'm} = \delta_{m'm}\delta\omega_m^\Omega 
    \label{eq:rot_splitting}
\end{align}
with 
\begin{equation}
\delta\omega_m^\Omega =m \int_V K_{n\ell m} (r,\theta) \left[ \Omega (r,\theta)-\Omega_\beta \right] \df{V} 
- m\Omega_\beta C_{n\ell} +  \eta Q_{2 \ell m}\,  \omega_{n\ell}^{(0)},
    \label{eq:rot_splitting2}
\end{equation}
where $\Omega(r,\theta)$ is the internal angular velocity in an inertial frame.
By construction, the angular velocity $\Omega_\beta$ of the frame $\mcR_\beta$ is $\Omega_\beta = \Omega( R,\beta)$, where $R$ is the radius of the star.
The first and second terms on the right-hand side of Eq. (\ref{eq:rot_splitting2}) describe the effect of differential rotation, where the functions $K_{n\ell m} (r,\theta)$ are the rotational sensitivity kernels \citep{Hansen1977} and $C_{n\ell}$ are the Ledoux constants \citep{1951ledoux} that account for the effect of the Coriolis force. 
The last term describes the quadrupole distortion of the star due to the centrifugal forces \citep[e.g.][]{Saio1981,Aerts2010} and is proportional to the ratio of the centrifugal to the gravitational forces at the surface $\eta=\Omega^2 R^3/(G M)$, where $M$ is the mass of the star and $G$ is the universal constant of gravity.
The term $Q_{2\ell m}$ accounts for the quadrupolar component of the centrifugal distorsion 
\begin{gather}
 Q_{2 \ell m}\simeq \frac{2/3 \int_{-1}^{1} P_2(x) \left[ P_\ell^{|m|}(x)\right]^2 \df{x}}
 {\int_{-1}^{1}  \left[ P_\ell^{|m|}(x)\right]^2 \df{x}} 
 = \frac{2\ell(\ell+1) - 6 m^2 }{3(2\ell+3) (2\ell-1)},
\end{gather}
where $P_\ell^m(x)$ are the associated Legendre functions and $P_2$ is the Legendre polynomial of second order.
The centrifugal term is very small in the case of slow rotators like the Sun \citep{1992dziembowski}, however it increases rapidly with rotation and it is not negligible anymore for stars rotating a few times faster than the Sun \citep[e.g.][]{2004gizon}.

\subsubsection{Frequency splittings due to the active region}

In this section we parametrize the effects  of the active region on the oscillation frequencies in the corotating frame $\wmcR_\beta$, where the active region is at the north pole.

\begin{figure}
 \includegraphics[width=\columnwidth]{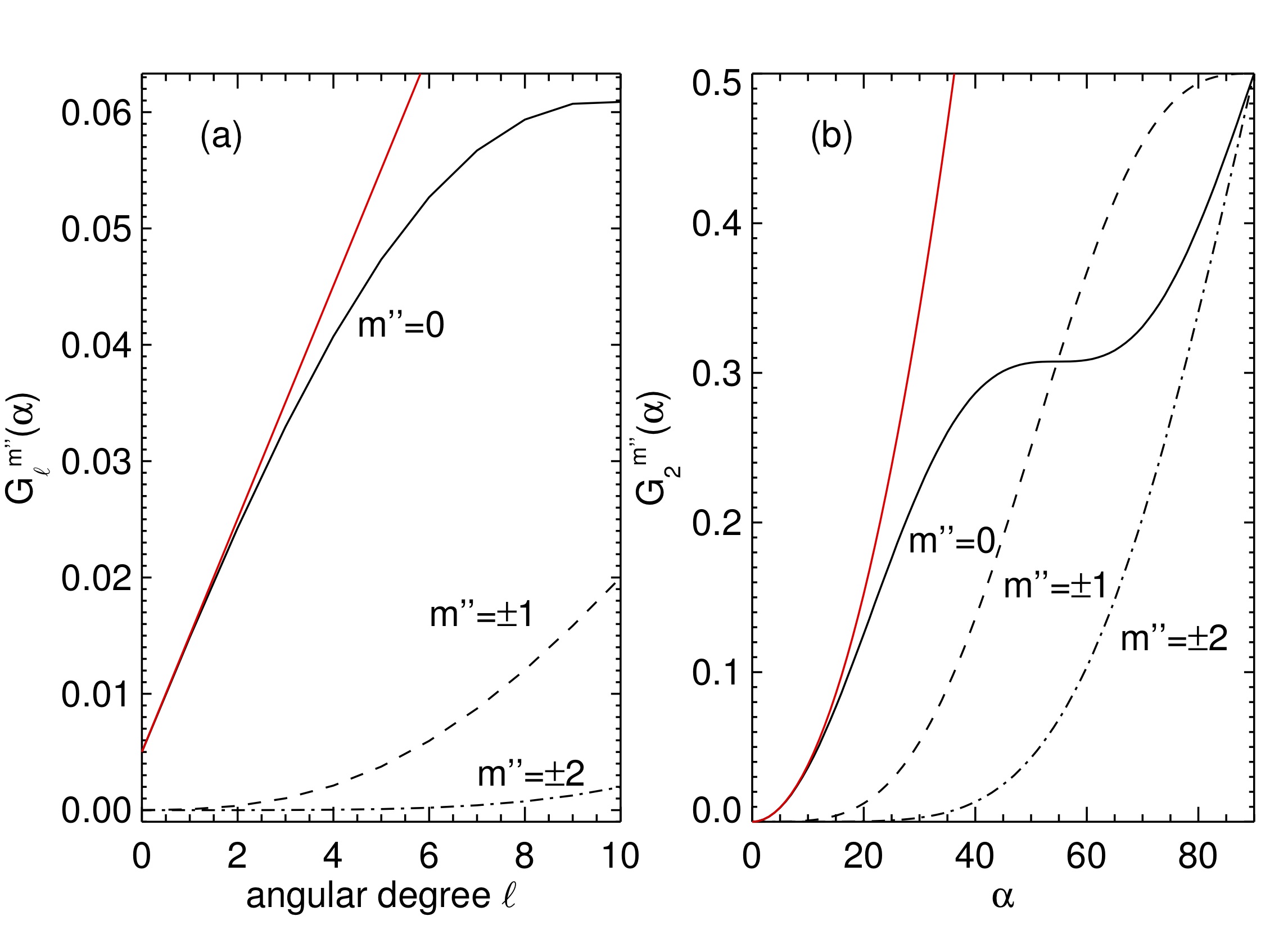}
 \caption{ Left panel: Geometrical weight factor $G_\ell^{m''}(\alpha)$ defined by Eq.~(\ref{eq:glm}) as a function of angular degree $\ell$ for $m=0$, $\pm1$, $\pm2$, at fixed $\alpha=8^\circ$. Right panel: $G_\ell^{m''}(\alpha)$ as a function of  $\alpha$ for $\ell=2$.
{ The red curves show the parabolic approximations for $G_\ell^0(\alpha)$ in the limit of small $\alpha$ (see Sec. \ref{sec:small_ar}, Eq. \ref{eq:Gl0_approx}).} 
 }
 \label{fig:glmalpha}
\end{figure}

Modeling the complex influence of surface magnetic fields on acoustic oscillations is challenging \citep[\eg,][]{Gizon2010,Schunker2013}.
Here we choose to drastically simplify the physics and to focus on the geometrical aspects of the problem. 
Assuming that the area of the active region covers a polar cap with $0\le \widehat \theta_\beta \le \alpha$ (see Fig.
\ref{fig:referenceframes}) and that the structure of the active region is separable in $r$ and $\widehat\theta_\beta$, we can parametrize the perturbation matrix as follows:
\begin{gather}
\label{eq:omat_mag_epsilon}
 \widehat O^{\mathrm{AR}}_{m''m'} = \delta_{m''m'} \delta\omega_{m'}^{\rm AR} = \delta_{m''m'} \omega_{n\ell}^{(0)} \,\varepsilon_{n\ell}\, G_\ell^{m''}(\alpha) \, ,
\end{gather}
where
\begin{gather}
\label{eq:glm}
  G_\ell^{m''}(\alpha) = \frac{(2\ell+1)(\ell-|m''|)!}{2(\ell+|m''|)!} 
      \int_{\cos\alpha}^1 \left [P_\ell^{|m''|}(x)\right ]^2 \df x
\end{gather}
is a geometrical weight factor that accounts for the surface coverage of the active region.
For small values of $\alpha$, $G_\ell^{m''}(\alpha)$ decreases fast as $|m''|$ increases (Fig. \ref{fig:glmalpha}). 
Since the value of $G_\ell^{m''}(\alpha)$ does not depend on the sign of $m''$, the eigenvalues of $\bO^{\rm AR}$ are degenerate in $|m|$ \citep[see also, \eg,][]{Kurtz1986}. 

The parameter $\varepsilon_{n\ell}$ is a measure of the relative magnitude of the active region perturbation and contains all the physics. 
From studies of local helioseismology \citep[\eg][]{Gizon2009, Moradi2010, Schunker2013}, it is known that the net effect of an active region is to increase the frequencies of acoustic modes, i.e. waves propagate faster in magnetic regions and thus $\varepsilon_{n\ell}$ is positive.
Since the active region introduces a perturbation that is strongly localized near the surface, the value of $\varepsilon_{n\ell}$ increases with radial order $n$. 
A proper calculation of the value of $\varepsilon_{n\ell}$ goes beyond the scope of the present study.
Instead, we parametrize the active region perturbation as an increase in sound speed near the surface.
Following \citet[][]{Papini2015}, we write
\begin{align}
\label{eq:epsilon_nldef}
 \varepsilon_{n\ell}= & \int_\mathrm{AR} \Delta c^2(r) \left( \frac{1}{r} \PD{\left(r^2\xi_{r,n\ell}(r)\right)}{r} - \ell(\ell+1)\xi_{h,n\ell} (r)\right)^2 \rho_0(r) \df{r} \nonumber\\
 &\times \left ( 2 \omega_{n\ell}^{(0)2} \int_V ||\bxi^{(0)}_{n\ell m}||^2\rho_0(r)\df{V}
 \right)^{-1} ,
\end{align}
where $\rho_0(r)$ is the density of the unperturbed stellar background and $\Delta c^2(r)$ is the radial change in the squared sound speed.
Here the integral is over the radial extent of the active region.  In Section~\ref{sec:AR_parameters} we will further specify the effective sound-speed perturbation.

Using Eq.~(\ref{eq:OtildeAR_toOAR}), the perturbation matrix elements  in frame $\mcR_\beta$ are
\begin{gather}
O^{\mathrm{AR}}_{m'm} = \omega_{n\ell}^{(0)} \varepsilon_{n\ell} 
\sum_{m''=-\ell}^\ell G_\ell^{m''}(\alpha) r^{(\ell)}_{m'' m'}(\beta) r^{(\ell)}_{m ''\, m}(\beta).
\end{gather}
We note that for $\ell\leq 3 $ and $\alpha \lesssim 30^\circ$, we have $G_\ell^{0}(\alpha) \gg  G_\ell^{m''}(\alpha)$, where $|m''| \neq 0$,
and therefore the dominant eigenvalue is 
$
O_{00}^\mathrm{AR}~=~\omega_{n\ell}^{(0)}\varepsilon_{n\ell} G_\ell^0(\alpha)
$, see Eq. (\ref{eq:omat_mag_epsilon}).

\subsection{Numerical setup for the linear problem}

We now introduce the internal rotation model and the active region parameters, used to illustrate the theory.
We consider a star with a rotation period of $8$~days, about one third the rotation period of the Sun. 
{ This choice of rotation period ensures that the azimuthal modes in a multiplet are well separated in frequency space.}
For the internal rotation profile, we take
\begin{gather}
  {\Omega(r,\theta)}/{2\pi} = 
  \left\{ 
  \begin{array}{ll}
  (1447 - 183 \cos^2\theta - 253 \cos^4\theta)
  \text{  nHz } &  r>0.7R_\odot , \\
 1447\text{  nHz } &  r< 0.7 R_\odot ,
 \end{array}
 \right.
 \label{eq.rotprofile}
\end{gather}
which is a scaled model of solar differential rotation as in \citet{2004gizon}. The centrifugal term has a significant effect. It shifts the $m=0$ mode and introduces an asymmetry in the shifts for positive and negative azimuthal orders $m$, with a maximum frequency shift of more than $100~\nHz$ in the case of a multiplet near $3$~mHz.
Therefore this term must be included when performing the analysis.

\label{sec:AR_parameters}

From observations of $p$-mode frequency changes during the solar cycle, \citet{Libbrecht1990} showed that the (positive) frequency shifts are almost independent of $\ell$ and increase with frequency, thus indicating that the effects of magnetic activity on acoustic oscillations are confined to the surface.
Assuming that the perturbation covers two pressure scale heights below the photosphere and setting $\Delta c^2/c^2 \simeq 10\%$ there, Eq. (\ref{eq:epsilon_nldef}) gives $\varepsilon_{n\ell}\simeq 0.003$  at frequencies near $3$~mHz.
The surface coverage of a stellar active region, as inferred from Doppler imaging, ranges from a percent up to $10\%$ \citep{Strassmeier2009}. Here we consider two different surface coverages of either $4\%$ or  $7\%$, corresponding to $\cos\alpha=0.92$ or $0.86$ ($\alpha\simeq23^\circ$ and $30^\circ $). 
Finally, we consider an active region at a colatitude of either $\beta=20^\circ$ (near the pole) or $80^\circ$ (near the equator).

In the following section we focus on two multiplets with $(\ell, n)=(1, 18)$ and $(2,18)$. For each of these modes we solve the eigenvalue problem (\ref{eq:eigen_problem}) by means of Jacobi's method \citep{numrecipes}. 
For the calculation of the unperturbed eigenmodes $(\omega_{n\ell}^{(0)},\bxi_{n\ell m}^{(0)})$ we use the ADIPLS software package \citep{Christensen-Dalsgaard2008} and Solar Model S  as the reference structure model \citep{JCD1996}.
The unperturbed frequencies $\omega_{n\ell}^{(0)}/2\pi$ of the dipole and quadrupole modes are $2695.40~\mu\Hz$ and $2756.95~\mu\Hz$ respectively.
{ We note that the choice of reference solar model is unimportant for the present study.}

\subsection{Power spectrum in the observer's frame: $(2\ell+1)^2$ peaks}
\label{sec:linear_PS}

Given particular values for $\alpha$, $\beta$, and $\varepsilon_{n\ell}$ the eigenvalue problem (\ref{eq:eigen_problem}) is fully specified and can be solved.
In this section we use the solutions (\ref{eq:xi_pert}) and (\ref{eq:freq_pert}) to build a synthetic power spectrum in the observer's frame, in order to relate the results to observations.

We need to find an expression that connects the eigenmodes to the observed intensity fluctuations.
For the sake of simplicity, we assume that the variation $I(\theta_\beta,\phi_\beta,t)$ induced by the acoustic oscillations in the emergent photospheric intensity is proportional to the Eulerian pressure perturbation $p$ of the acoustic wavefield \citep[\eg][]{Toutain1993}, as measured  at the stellar surface $r=R$. 
The pressure perturbation $p$ is related to the wavefield  displacement $\bxi$ through the linearized adiabatic equation
\begin{gather*}
p = -\rho_0 c_0^2 \nabla \cdot \bxi - \bxi\cdot \nabla P_0,
\end{gather*}
where $c_0$ and $P_0$ are respectively the sound speed and the pressure of the unperturbed stellar model.
The pressure perturbation $p_{n\ell M}(r,\theta_\beta,\phi_\beta)$ of the mode $M$ is
\begin{equation}
p_{n\ell M} = -\rho_0 c_0^2 \nabla \cdot \bxi_{n\ell M} - \bxi_{n\ell M}\cdot \nabla P_0,
\end{equation}
where $\bxi_{n\ell M}$ is given by Eq. (\ref{eq:xi_pert}).
To  leading order, we have
\begin{gather}
p_{n\ell M}({\br_\beta})= \sum_{m=-\ell}^\ell A_m^M p_{n\ell m}^{(0)}(\br_\beta)
\end{gather}
with 
\begin{gather*}
p_m^{(0)}(R,\theta_\beta,\phi_\beta) = 
 -\rho_0 c_0^2 \nabla \cdot \bxi_m^{(0)} - \bxi_m^{(0)}\cdot \nabla P_0 ,
\end{gather*}
where we dropped the subscripts $n\ell$.
Acoustic oscillations in stars are stochastically excited and damped by turbulent convection, therefore $I(\theta_\beta,\phi_\beta,t)$ is a realization of a random process. 
Since  the perturbation is steady in the frame $\mcR_\beta$, this random process is stationary in that frame.
An expression for the intensity fluctuations $I(\theta_\beta,\phi_\beta,\omega)$ in  Fourier space with the required  statistical properties is
\begin{align}
I(\theta_\beta,\phi_\beta,\omega) 
& \propto \sum_{M=-\ell}^\ell p_M(R,\theta_\beta,\phi_\beta)
L_M^{1/2}(\omega) \mcN_M(\omega) \nonumber \\
 & \propto \sum_{M=-\ell}^\ell \sum_{m=-\ell}^\ell A_m^M \Ylm(\theta_\beta,\phi_\beta)
L_M^{1/2}(\omega)
    \mcN_M(\omega) ,
\label{eq:Iomega_corot}
\end{align}
where the $\mcN_M(\omega)$ are independent complex Gaussian random variables, with zero mean and unit variance:
\begin{gather}
\label{eq:covNMNMp}
  E \left [ {\mcN_{M'}}^*(\omega')\mcN_M(\omega)\right ] = \delta_{M'M} \delta_{\omega'\omega}.
\end{gather}
In Equation (\ref{eq:Iomega_corot}) and (\ref{eq:covNMNMp}) we only consider the positive-frequency part of the spectrum ($\omega$ and $\omega'>0$) since $I(\theta_\beta,\phi_\beta,t)$ is real. The negative-frequency part is related to the positive part by $I(\theta_\beta,\phi_\beta,-\omega)= I^*(\theta_\beta,\phi_\beta,\omega)$.
The function $L_M(\omega)$ is a Lorentzian  
\begin{gather}
L_M(\omega)=\left [ 1+\left (\frac{\omega-\omega_M}{\Gamma/2}  \right )^2 \right ]^{-1},
\end{gather}
appropriate for describing the power spectrum of an exponentially damped oscillator  with full width at half maximum (FWHM) $\Gamma$ \citep[see ,e.g.,][]{Anderson1990}.

We transform Eq. (\ref{eq:Iomega_corot}) back to the time domain by inverse Fourier transformation to obtain $I(\theta_\beta,\phi_\beta,t)$.
The intensity $I(\theta,\phi,t)$ as seen by the observer in the inertial frame $\mcR$ is obtained by applying a passive rotation of Euler angles $(0,-i,\Omega_\beta t)$ to
express $\Ylm(\theta_\beta,\phi_\beta)$ in terms of $\theta$ and $\phi$ \citep{1959messiah}:   
\begin{gather}
Y_\ell^{m}(\theta_\beta,\phi_\beta) = 
e^{-\mathrm{i} m\Omega_\beta t}
\sum_{m'=-\ell}^{\ell} Y_\ell^{m'}(\theta,\phi) r_{m'm}^{(\ell)}(-i) .
\end{gather}
In the frequency domain, the intensity fluctuations become
\begin{equation}
I (\theta,\phi,\omega)  = \sum_{M=-\ell}^\ell \sum_{m=-\ell}^\ell\sum_{m'=-\ell}^\ell 
A_m^M Y_\ell^{m'}(\theta,\phi) r_{m\,m'}^{(\ell)}(i) \;  L_M^{1/2}(\omega - m\Omega_\beta)
    \mcN_M(\omega- m\Omega_\beta) ,
\label{eq:I_thetaphiomega}
\end{equation}
where we used the property $r_{m'm}^{(\ell)}(-i)=r_{m\,m'}^{(\ell)}(i)$.
Since $L_M(\omega - m\Omega_\beta)$ peaks at frequency $\omega=\omega_M + m\Omega_\beta$, the intensity spectrum observed in the inertial frame has $(2\ell+1)^2$ peaks, corresponding to all combinations of $m$ and $M$.

To obtain the full-disk integrated intensity fluctuations  $I_\text{obs}(\omega)$ we perform an integration over the visible disk of the star:
\begin{gather*}
 I_\text{obs}(\omega) = 
\int_0^{2\pi} \df\phi \int_0^{\pi/2} \df\theta \;
 I(\theta,\phi,\omega)  W(\theta)\cos{\theta}\sin\theta \,,
\end{gather*}
where $W(\theta)$ is the limb-darkening function. 
The components with $m'\ne 0$ vanish upon integration over $\phi$, thus 
\begin{gather}
\label{eq:Iobs_final}
I_\text{obs} (\omega) 
 =  \sum_{M=-\ell}^\ell \sum_{m=-\ell}^\ell
B_m^M L_M^{1/2}(\omega-m\Omega_\beta)
    \mcN_M(\omega- m\Omega_\beta) ,
\end{gather}
with
$B_m^M =  A_m^M \,V_\ell \; r_{m0}^{(\ell)}(i)$ and
\begin{gather}
\label{eq:V_l}
V_\ell=
\int_0^{2\pi} \df\phi \int_0^{\pi/2} \df\theta\; 
 Y_\ell^0(\theta,\phi)  W(\theta)\cos{\theta}\sin\theta .
 \end{gather}
The matrix elements $r_{m0}^{(\ell)}(i)$ are written explicitly in terms of the associated Legendre polynomials \citep{1959messiah}:
\begin{gather}
\label{eq:r0lm}
 {r}_{m0}^{(\ell)}(i) =
 (-1)^m \sqrt{\frac{(\ell-m)!}{(\ell+m)!}} P_\ell^m(\cos i).
\end{gather}

\begin{figure*}
\includegraphics[width=20cm]{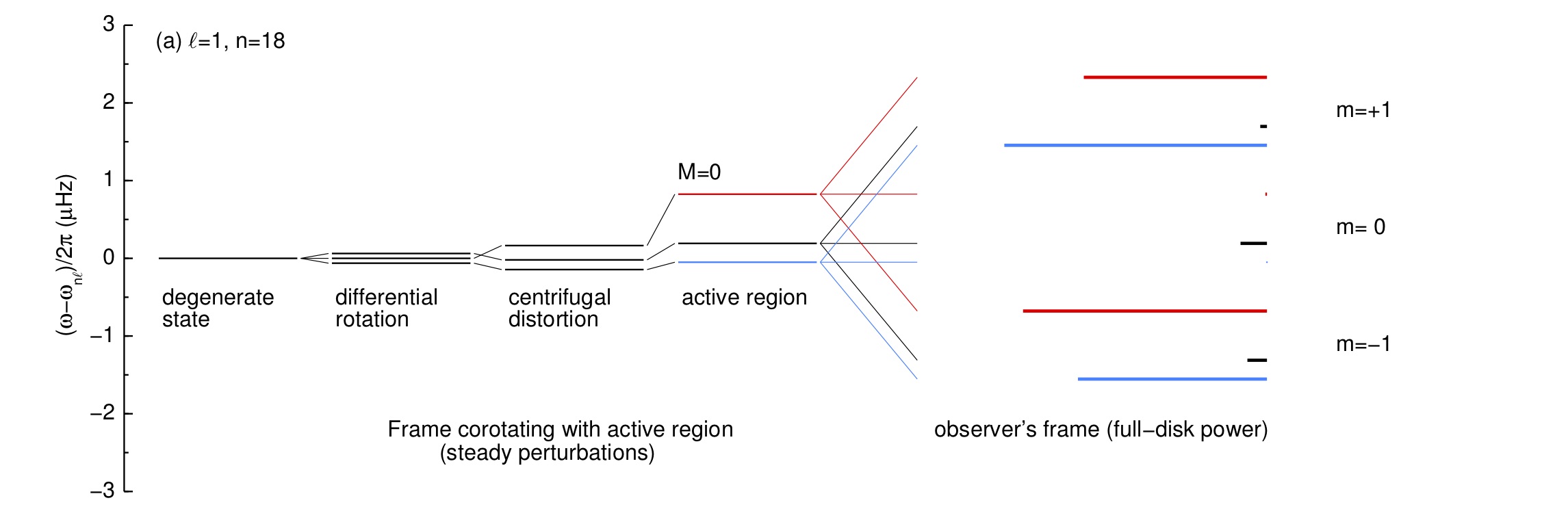}
\includegraphics[width=20cm]{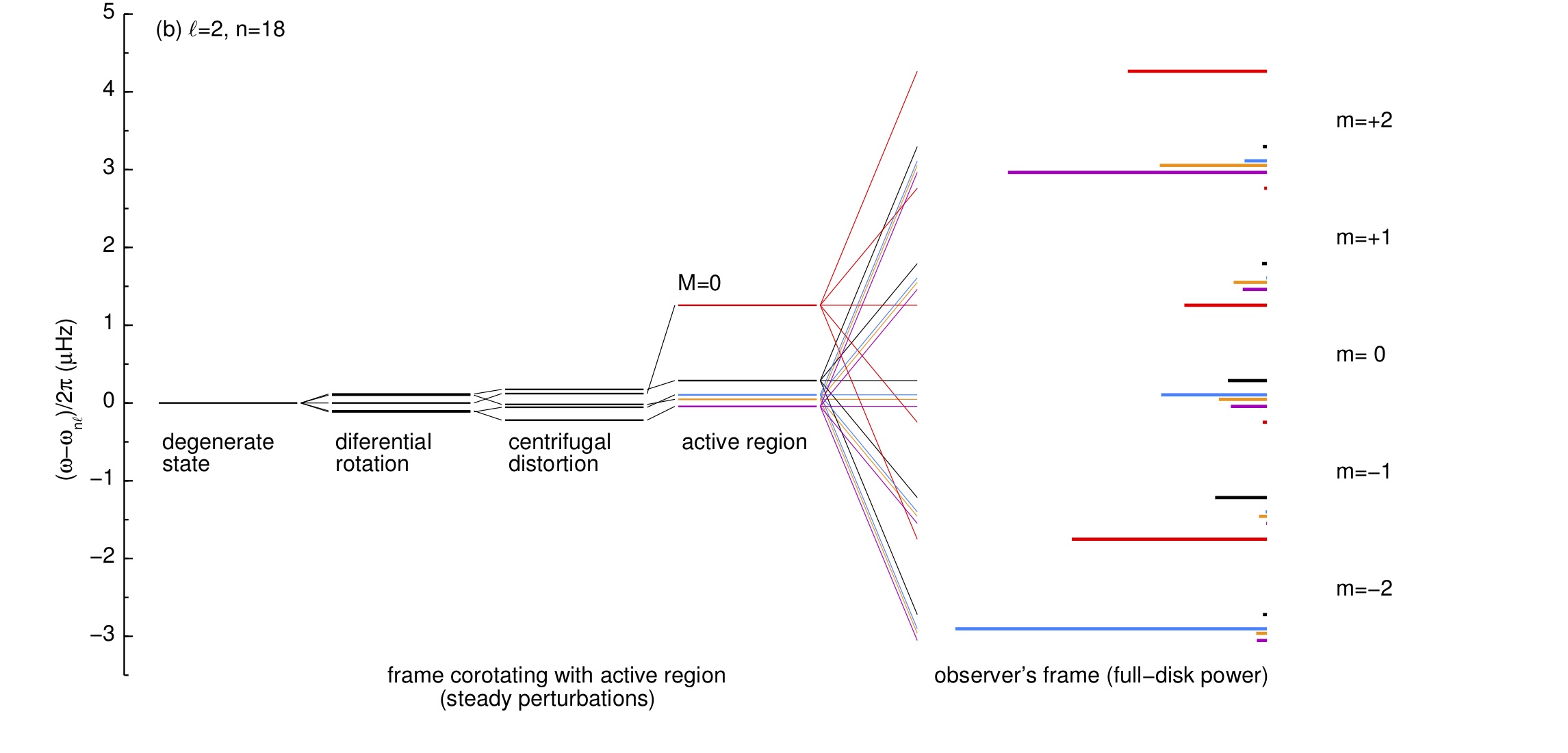}
 \caption{ Left side of the diagram: Perturbations to the mode eigenfrequencies in the frame that is corotating with the active region,  for the multiplets with $(\ell,n)=(1,18)$ (top) and with $(\ell,n)=(2,18)$ (bottom). The star is Sun-like and rotates with a scaled solar differential rotation profile (rotation period is approximately $8~\text{days}$). The active region perturbation is specified by  $\varepsilon_{n\ell}=0.003$, $\alpha=23^\circ$, and $\beta=80^\circ$. The rotational frequency of the active region at colatitude $\beta$ is $\Omega_\beta/2\pi$ is $1.504~\mu\Hz$.
The frequency of the mode $M=0$  is the most shifted. Right side of the diagram: The  $(2\ell+1)^2$ peaks of the power spectrum as seen in the observer's frame, for an inclination angle $i=80^\circ$. For each $m$, the $M$-components are identified with different colors: red for $M=0$, black for $M=1$, blue for $M=-1$, orange for $M=2$, and pink for $M=-2$.  
 The peaks with the same colors are statistically correlated to each other (according to  Eq.~\ref{eq:PS_correlation}).
 }
 \label{fig:degenerate2obs}
\end{figure*}

A realization of the power spectrum is given by   
\begin{align}
P(\omega) &= |I_\text{obs}(\omega)|^2 \nonumber\\
 &=  \left |\sum_{M=-\ell}^\ell \sum_{m=-\ell}^\ell
     B_m^M L_M^{1/2}(\omega-m\Omega_\beta)
    \mcN_M(\omega- m\Omega_\beta) \right |^2,
\label{eq:powerspectrum_realization}
\end{align}
which depends on $2\ell+1$ independent realizations of complex Gaussian random variables. The expectation value of $ P(\omega)$ is
\begin{gather}
\label{eq:powerspectrum}
\mc{P}(\omega) = E\left [P(\omega)\right ]=  \sum_{m=-\ell}^\ell \sum_{M=-\ell}^\ell
 P_m^M L_M(\omega - m\Omega_\beta).
\end{gather}
As mentioned earlier, the power spectrum displays $(2\ell+1)^2$ Lorentzian peaks, centered at frequencies 
\begin{gather}
\label{eq:observed_frequencies}
 \omega_m^M := \omega_M + m \Omega_\beta = \omega_{n\ell}^{(0)}+\delta\omega_M + m \Omega_\beta  ,
\end{gather}
 with peak power
\begin{gather}
\label{eq:observed_amplitude}
 P_m^M := ({B_m^M})^2 = \frac{(\ell-|m|)!}{(\ell+|m|)!} \left [V_\ell A_m^M P_\ell^{|m|}(\cos i) \right ]^2.
\end{gather}
Example power spectra for a dipole and a quadrupole multiplet are shown in Fig. \ref{fig:degenerate2obs}.
The frequencies and amplitudes of the $(2\ell+1)^2$ peaks are obtained from Eqs.  (\ref{eq:observed_frequencies}) and (\ref{eq:observed_amplitude}),  using the limb-darkening function quoted by \citet{Pierce2000} to calculate $V_\ell$.

Figure \ref{fig:degenerate2obs} also displays the different contributions to the frequency splittings due to rotation and to the active region perturbation in the corotating frame $\mcR_\beta$.
For both multiplets, the $M=0$ peaks are shifted by the largest amount, they are the most affected by the AR perturbation and in the frame $\wmcR_\beta$  \citep[see also][]{Papini2015}.
This feature, which arises from geometrical considerations only,  is preserved in the spectrum as seen in the observer's frame, where  the $M=0$ peaks are clearly visible.
With increasing AR surface coverage the frequency shifts of the $M\neq 0$ modes increase and the peaks get less clustered.

\section{Results}

\subsection{Dipole and quadrupole power spectra}
In this section we describe the changes  imprinted by a large active region in the spectrum of two multiplets with $(\ell,n)=(1,18)$ and $(2,18)$, 
for a star rotating with a period of $8~\text{days}$.

\begin{figure}
\begin{center}
 \includegraphics[width=0.9\columnwidth]{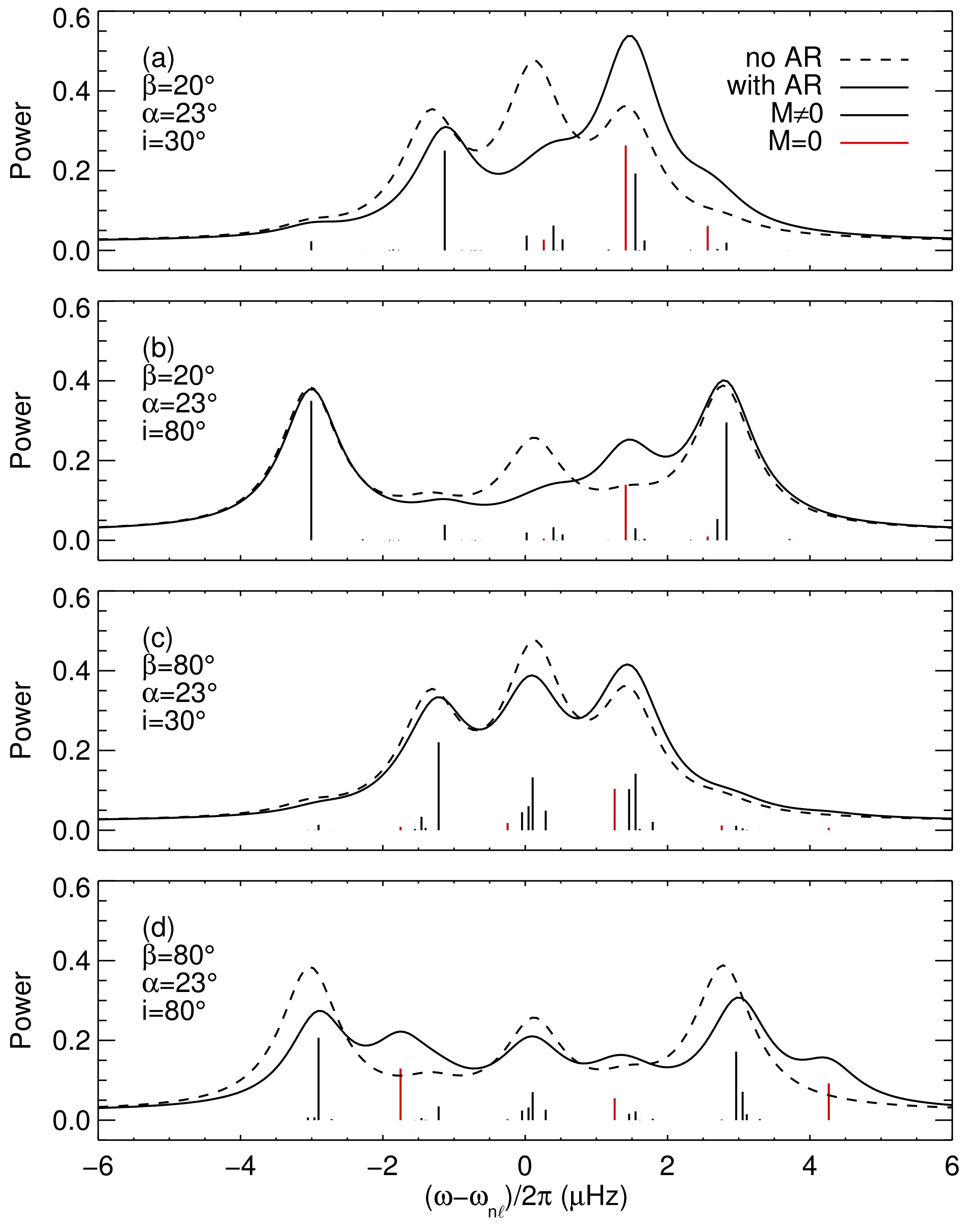}
 \end{center}
 \caption{Oscillation power spectra for the $(\ell,n)= (2,18)$ multiplet observed at two inclination angles $i=30^\circ$ (panels a and c) and $80^\circ$ (panels b and d),
 for a star with a rotation period of $8~\text{days}$ and for an active region with $\varepsilon_{n\ell}=0.003$, $\beta=20^\circ$ (panels a and b) or $80^\circ$ (panels c and d), and for a surface coverage with $\alpha=23^\circ$. {The power spectra are normalized with respect to $V_2$ (Eq. \ref{eq:V_l}).}  The vertical line segments show the theoretical frequencies and amplitudes for $M=\pm1,\pm 2$ modes (black) and the $M=0$ mode (red). 
 The envelopes of the power spectra (solid black curves) are obtained by summing over Lorentzians with widths of $1~\mu\Hz$. The dashed black curve shows the envelope of the pure rotational power spectrum, which includes the centrifugal distortion (Eq. \ref{eq:rot_splitting}).
 }
 \label{fig:l2n18spectrum}
\end{figure}
Figure \ref{fig:l2n18spectrum} shows the results for the quadrupole multiplet, for four combinations of the values of $\alpha$ and $\beta$ selected in Sect.~\ref{sec:AR_parameters}: the observed power spectra are plotted for two angles of observation, $i=30^\circ\text{ and}~80^\circ$, and are 
normalized with respect to $V_2$, \ie~with respect to the $m=0$ peak of the pure rotational spectrum seen with $i=0$.
The corresponding theoretical Lorentzian envelope (solid line) has been calculated 
from Eq. (\ref{eq:powerspectrum}), by setting a value for the FWHM of $\Gamma/2\pi=1~\mu\Hz$, typical for this multiplet in the Sun \citep[see, \eg,][]{Chaplin2005}.
Due to the finite lifetime of the modes of oscillation, it is clear that is not possible to resolve all the $(2\ell+1)^2$ peaks, and an observer would identify not many more than $(2\ell+1)$ peaks in a multiplet.
In the cases shown here, it is possible to identify from 5 to 6 peaks for $i=80^\circ$, the additional peak coming from the uppermost shifted $m=2,M=0$ peak. 
We note that the Lorentzian envelope displays an asymmetric profile.
Because of their large shifts in frequency, the $M=0$ peaks blend with peaks from different $m$-quintuplets. Blending increases with activity level. Figure \ref{fig:l2n18spectrum}a shows a case for which the $(M,m)=(0,0)$ and  $(M,m)=(1,1)$ peaks have close frequencies and comparable amplitudes, they contribute equally to a single peak in the power spectrum.
 
\begin{figure}
\includegraphics[width=0.97\columnwidth]{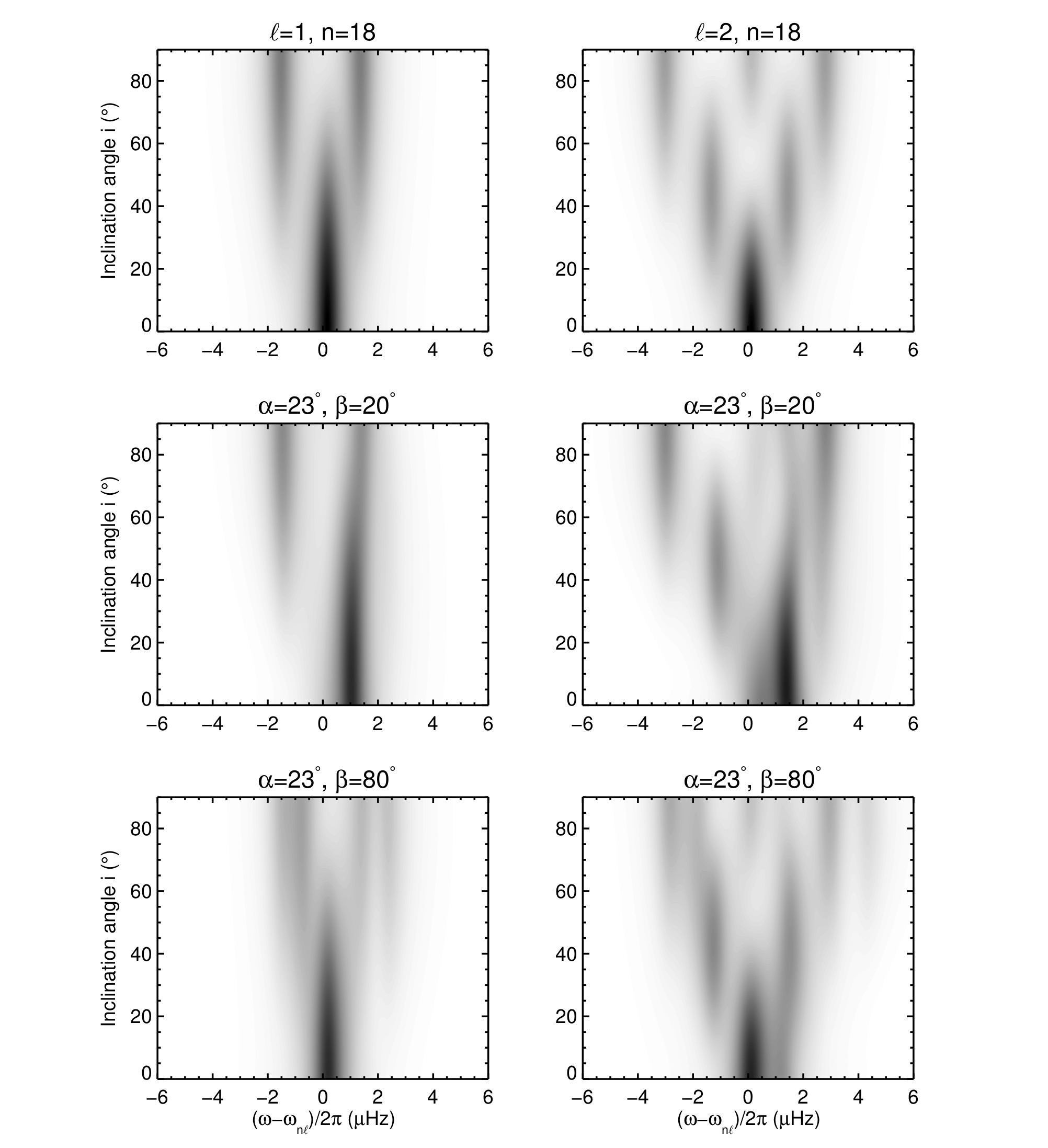}
 \caption{
 Expectation value of power spectra of oscillation,  as functions of inclination angle $i$.
 The left panels are for the dipole multiplet $\ell=1,n=18$, and the right panels for the quadrupole multiplet $\ell=2,n=18$.
 The top panels are for the pure-rotation case, the middle panels are for an active region at co-latitude $\beta=20^\circ$, and the bottom panels for  $\beta=80^\circ$.
The active region parameters are  $\varepsilon_{n\ell}=0.003$, $\alpha=23^\circ$, and the stellar rotation period is $8$~days.
 }
 \label{fig:contours}
\end{figure}

\begin{figure}
\includegraphics[width=1.0\columnwidth]{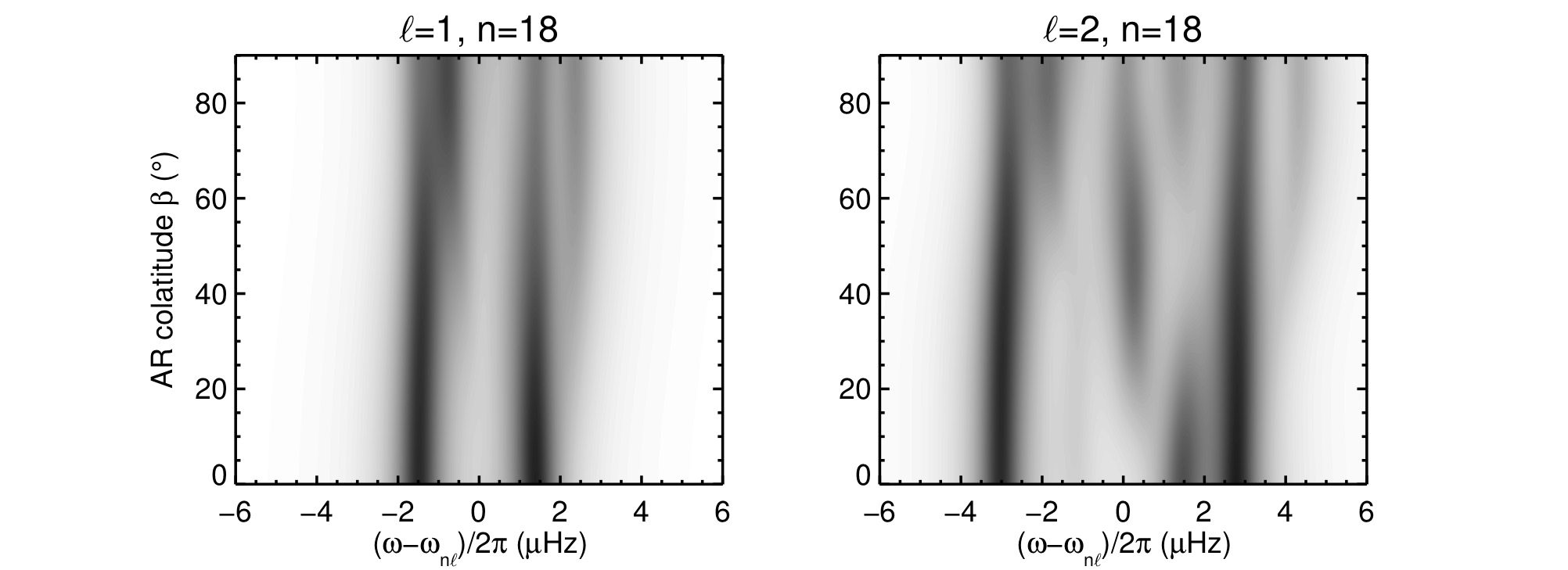}
 \caption{Expectation value of power spectra of oscillation, as functions of active-region colatitude $\beta$, at fixed inclination angle $i=80^\circ$. The other physical parameters are the same as in Fig.~\ref{fig:contours}.
 }
 \label{fig:contours_beta}
\end{figure}

The  envelope of the power spectrum  is very sensitive to the latitudinal position of the AR and to the inclination angle: in Fig. \ref{fig:l2n18spectrum}c the power spectrum is near the standard rotationally-split spectrum, while the same configuration observed from a different inclination angle (Fig. \ref{fig:l2n18spectrum}d) shows a more asymmetric profile with additional peaks. 
{This is better seen in Fig. \ref{fig:contours}, which shows contours of the acoustic power as a function of inclination angle, for both the $\ell=1$ and $\ell=2$ multiplets, for the same active region parameters as in Fig. \ref{fig:l2n18spectrum}. For an active region at high latitude (middle panels of Fig. \ref{fig:contours}), the central peak  shows a significant shift and overlaps with the $m=1$ peak. 
For a low-latitude active region (bottom panels) the envelope of the power spectrum displays more than $2\ell+1$ peaks. A distinct feature is the presence of  two peaks  instead of one when observing an $\ell=2$ multiplet at zero inclination angle.}
{The sensitivity of the spectrum to the colatitude of the AR, shown in Fig \ref{fig:contours_beta}, is due in part to the variation with $\beta$ of the non-diagonal elements of the rotation matrix ${\bR}^{(\ell)}$. }

\begin{figure}
\includegraphics[width=1.0\columnwidth]{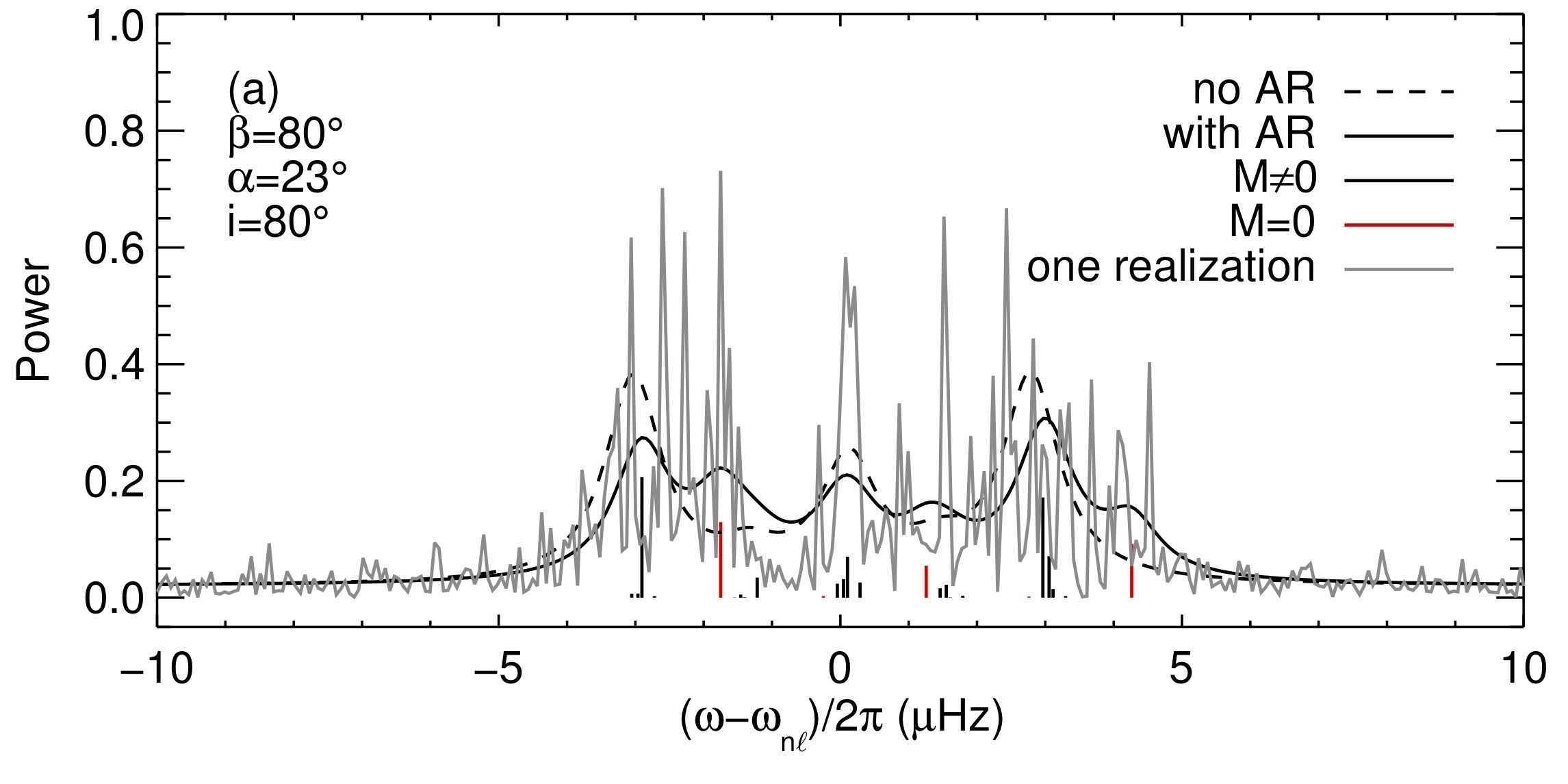}
 \caption{A realization of the power spectrum of an $\ell=2$ multiplet (gray) and its expectation value (black solid curve, also shown in  Fig.~\ref{fig:l2n18spectrum}d). The observation duration is 6 months and the signal-to-noise ratio is 50.  The dashed black curve shows the expectation value of the pure rotational spectrum.
 }
 \label{fig:noise_1}
\end{figure}

An observed power spectrum is, of course, much more difficult to interpret than its expectation value. The power spectrum in Fig.~\ref{fig:noise_1} includes realization noise due to the stochastic nature of stellar oscillations and to additional shot noise. At each frequency the observed power is a realization of an exponential distribution (a chi-squared with two degrees of freedom) with standard deviation and mean equal to the expectation value of the power spectrum. Realization noise considerably degrades the spectrum, however in some cases it is still possible to distinguish between the pure rotational spectrum and a spectrum with the active region.

\subsection{Asymptotics}
\subsubsection{Limit of small latitudinal differential rotation}
\label{sec:small_diff_rot}
The examples shown so far suggest that the power spectrum of a multiplet may be approximated by the sum of $2(2\ell+1)$ peaks, for two reasons.
First, in the corotating frame the splitting due to rotation is small compared to the shift induced by the AR perturbation (see, \eg, Fig. \ref{fig:degenerate2obs}).
Second, the AR perturbation induces a shift that is largest for the $M=0$ mode. Therefore, for each $m$, all the $M\neq0$ peaks are clustered near the pure rotational frequencies and appear as a single peak, while the $M=0$ peaks are well separated. 

Here we wish to find an approximation for the power spectrum of a multiplet in terms of $2(2\ell+1)$ Lorentzians only. 
In the following we assume the rotation perturbation to be small compared to the active region perturbation, and seek an approximate solution to the eigenvalue problem (\ref{eq:eigen_problem}). 

It is convenient to solve the eigenvalue problem in the frame where the dominant perturbation is diagonal, \ie, the frame $\wmcR_\beta$ of the active region. The following results are similar to Sec. 19.5 of \citet[][]{Unno1979}. 
We rewrite the elements of the full matrix $\widehat{\bO}$ in $\wmcR_\beta$ as
\begin{gather}
\widehat{O}_{m''m'} = \delta_{m''m'}\delta\omega_{m'}^{\rm AR} +
\sum_{m=-\ell}^{\ell}
r_{m''m}^{(\ell)} (\beta) r_{m'm}^{(\ell)} (\beta)
\delta\omega_{m}^\Omega ,
\end{gather}
where the term with the sum in the right hand side correspond to $\widehat O_{m''m'}^\Omega $.
We look for solutions of the form
\begin{gather}
\delta\omega_M = \delta\omega_{M}^{\rm AR} + \delta\omega_M^{(1)}
\end{gather}
and 
\begin{gather}
\widehat\bA{}^M = \widehat\bA{}^{M,\rm AR} + \widehat\bA{}^{M,(1)},
\end{gather}
where $\left (\delta\omega_M^{(1)},\widehat\bA{}^{M,(1)}\right )$ are the perturbations to the (partially degenerate) eigenvalues $\delta\omega_{M}^\mathrm{AR}$ and eigenvectors $\widehat\bA{}^{M,\rm AR}$ of $\widehat\bO{}^\mathrm{AR}$, with $\widehat{A}_m^{M,\mathrm{AR}}=\delta_{mM}$. 
The eigenvector perturbation is orthogonal to the reference eigenspace  \citep[see, \eg,][p. 687]{1959messiah}, 
\beq
\widehat{\bA}^{M,(1)} \cdot \widehat{\bA}^{\pm M,\mathrm{AR}}=0,
\label{eq:orthoA'A}
\eeq
which gives 
$
\widehat A_{\pm M}^{M,(1)} = 0.
$
The eigenvectors  $\bA^M$ in the frame $\mcR_\beta$ are given by
\begin{gather}
\bA^M = \bR^{(\ell)} \widehat\bA{}^M 
\end{gather}
via the rotation matrix $\bR^{(\ell)}$ (see Eq.~\ref{eq:rotylmabg}).

To first order , the eigenvalue problem (Eq.~\ref{eq:eigen_problem})  becomes
\begin{gather}
 \left ( \widehat{\bO}^\Omega - \delta\omega_M^{(1)} \bI \right ) \widehat\bA{}^{M,\mathrm{AR}} + \left ( \widehat\bO^\text{AR} -\delta\omega_{M}^\mathrm{AR} \bI \right )\widehat\bA{}^{M,(1)} = 0,
\label{eq:eigenproblem_small}
\end{gather}
where $\bI$ is the identity matrix. 
To calculate $\widehat\bA{}^{M,(1)}$ and $\delta\omega_M^{(1)}$ we multiply the above equation on the left by the transpose of $\widehat\bA{}^{m',\text{AR}}$ to obtain:
\begin{gather}
\widehat O_{m'M}^\Omega  - \delta\omega_M^{(1)} \delta_{m'M} + 
\left (\delta\omega_{m'}^\mathrm{AR} -\delta\omega_{M}^\mathrm{AR} \right ) \widehat A_{m'}^{M,(1)}  = 0  ,  \quad   -\ell \le m' \le \ell.
\label{eq:eigenproblem_small_component}
\end{gather}
We find the perturbed eigenvalues by setting $m'=M$ in the above equation
\begin{equation}
\delta\omega_M^{(1)} =  \sum_{m=-\ell}^{\ell}
[r_{Mm}^{(\ell)} (\beta) ]^2 
\delta\omega_{m}^\Omega .
\end{equation}
The non-zero elements of $\bA^{M,(1)}$ are obtained from the $m'\neq \pm M$ components of Eq. (\ref{eq:eigenproblem_small_component}):
\begin{gather}
\widehat A_{m'}^{M,(1)} = \frac{\widehat O_{m'M}^\Omega}{\delta\omega_{M}^\mathrm{AR}-\delta\omega_{m'}^\mathrm{AR}} \quad \text{ for } m'\neq \pm M .
\end{gather}
The explicit expressions for the eigenvalues and eigenvectors of the combined perturbations are
\begin{align}
\label{eq:shifts_small_rotation}
\delta\omega_M = & \,\delta\omega_{M}^\mathrm{AR} + \sum_{m=-\ell}^{\ell}
[r_{Mm}^{(\ell)} (\beta) ]^2 
\delta\omega_{m}^\Omega 
\end{align}
and
\begin{gather}
\label{eq:amplitudes_small_rotation}
\widehat A_{m'}^M = \left \{
\begin{aligned}
&1 & \text{for } m'&=M ,\\
&0 & \text{for } m'&=-M ,\\
& \ffrac{\sum_{m=-\ell}^{\ell}
r_{m'm}^{(\ell)} (\beta) r_{Mm}^{(\ell)} (\beta)
\delta\omega_{m}^\Omega}
{\delta\omega_{M}^\mathrm{AR}-\delta\omega_{m'}^\mathrm{AM}}
& \text{for } m' &\neq M .
\end{aligned}
\right .
\end{gather}
The explicit expression for the amplitudes $A_m^M$ in the frame $\mcR_\beta$ is then
\begin{align}
A_m^M  = &\sum_{m'=-\ell}^\ell  \widehat A_{m'}^M r_{m'm}^{(\ell)}(\beta) \nonumber\\
= &r_{Mm}^{(\ell)}(\beta) 
 + \sum_{m'=-\ell,m'\neq \pm M}^{\ell}
\frac{r_{m'm}^{(\ell)}(\beta)}{\delta\omega_{M}^\mathrm{AR}-\delta\omega_{m'}^\mathrm{AR}}
\sum_{m''=-\ell}^{\ell}
r_{m'm''}^{(\ell)} (\beta) r_{Mm''}^{(\ell)} (\beta)
\delta\omega_{m''}^\Omega .
\label{eq:amplitudes_asymptotics}
\end{align}

\subsubsection{Neglecting latitudinal differential rotation}

If we neglect differential rotation then Eq. (\ref{eq:rot_splitting2}) reduces to
\beq 
\delta\omega_m^\Omega = -m\Omega_\beta C_{n\ell} +  \eta Q_{2\ell m} \, \omega_{n\ell}^{(0)}.
\label{eq:domega_nodiff_rot}
\eeq
Then, by using the identities \citep{Unno1979,1990gough}
\begin{equation}
	\sum_{m''=-\ell}^\ell 
    \left [r_{m'm''}^{(\ell)}(\beta) \right ]^2 m'' = m' \cos\beta 
\end{equation}
and
\begin{equation}
  \sum_{m''=-\ell}^\ell \left [r_{m'm''}^{(\ell)}(\beta) \right ]^2
    Q_{2 \ell m''}=P_2(\cos\beta) Q_{2 \ell m'},
\end{equation}
the frequency shifts (Eq.~\ref{eq:shifts_small_rotation}) simplify to
\begin{align}
\delta\omega_M = \delta\omega_{M}^\mathrm{AR} 
- M C_{n\ell} \Omega_\beta \cos\beta 
+ \eta P_2(\cos\beta) Q_{2 \ell M} \, \omega_{n\ell}^{(0)} .
 \label{eq:lin_dom_no_diff}
\end{align}
Note that for moderately fast rotating stars the Coriolis term  in the above equation is much smaller than centrifugal distortion term \citep[e.g.,][]{2004gizon}.

Due to the clustering of the $M\neq0$ peaks, the power spectrum in the observer's frame can be approximately modeled by $2(2\ell+1)$ Lorentzians, some of which may overlap. Half of these correspond to the peaks with $M=0$. The remaining $2\ell+1$ (approximate) Lorentzians are obtained by summing over the $M\neq0$ peaks; their mean frequency shifts are given by a power weighted average of the $M\neq0$ frequency shifts $\delta\omega_M^m = \delta\omega_M +m\Omega_\beta$. We denote by $\langle \delta\omega\rangle_m $ this average:
\begin{align}
{\langle \delta\omega\rangle}_m = 
(\Sigma_m )^{-1}
\left( \sum_{1\leq |M| \leq \ell} ({A_m^M})^2 \delta\omega_M  \right)
+m\Omega_\beta,
\label{eq:linear_freq_shift_Mneq0}
\end{align}
with 
\begin{align}
\Sigma_m = \sum_{1\leq |M| \leq \ell} {(A_m^M)}^2.
\end{align}
The corresponding averaged power amplitudes $\langle P\rangle_m $ are (see Eq.~\ref{eq:observed_amplitude})
\begin{gather}
 \langle P \rangle_m = \frac{(\ell-|m|)!}{(\ell+|m|)!} \left [V_\ell P_\ell^{|m|}(\cos i) \right ]^2 
 \Sigma_m.
 \label{eq:linear_amplitude_shift_Mneq0}
\end{gather}

\begin{figure}
\includegraphics[width=\columnwidth]{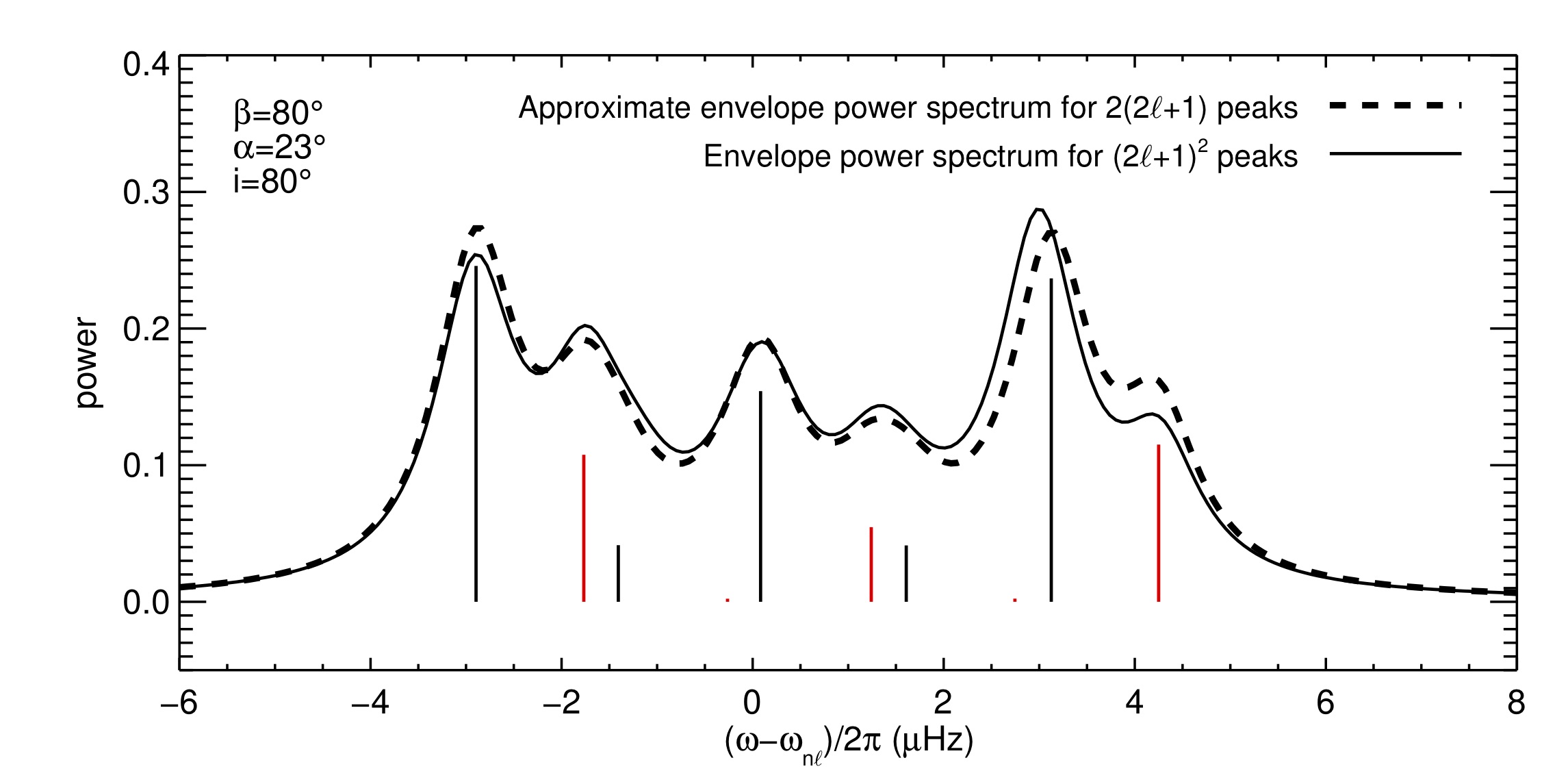}
\caption{Acoustic power spectrum for the same case as in Fig. \ref{fig:l2n18spectrum}d, (black solid line) and the power spectrum resulting from neglecting differential rotation and by assuming a small rotation perturbation with respect to the AR perturbation (black dashed line), as calculated from Eqs. (\ref{eq:amplitudes_asymptotics}, \ref{eq:domega_nodiff_rot}, \ref{eq:lin_dom_no_diff}).
Vertical red lines denote the $M=0$ peaks, and black lines the averaged $M\neq 0$ peaks from Eqs. (\ref{eq:linear_freq_shift_Mneq0}, \ref{eq:linear_amplitude_shift_Mneq0}). }
\label{fig:spectrum_small_diff_rot}
\end{figure}

\begin{figure}[t]
\includegraphics[width=\columnwidth]{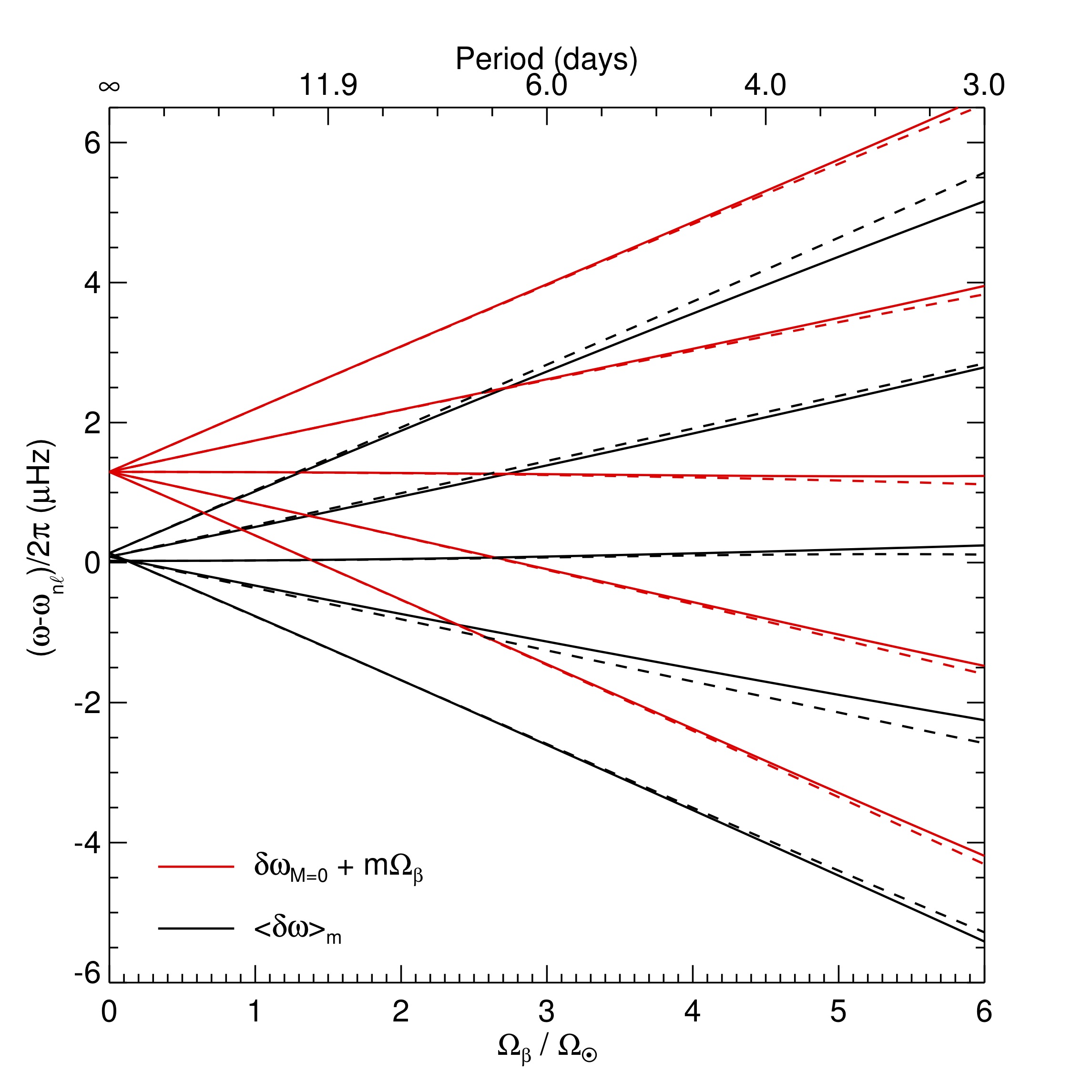}
\caption{
Black lines: Averaged frequency shifts $\langle \delta\omega\rangle_m$ vs. stellar rotation rate, as given by Eq. (\ref{eq:linear_freq_shift_Mneq0}) and calculated using Eq. (\ref{eq:eigen_problem}) (solid lines) and using the approximations of Eqs. (\ref{eq:amplitudes_asymptotics}, \ref{eq:domega_nodiff_rot}, \ref{eq:lin_dom_no_diff}) (dashed lines).
Red lines: Approximate frequency shifts $\delta\omega_{M=0}+m\Omega_\beta$ of the $M=~0$ peaks as given by Eq. (\ref{eq:lin_dom_no_diff}) (dashed lines) and first-order exact shifts (solid lines).
}
\label{fig:linear_qnl_AR}
\end{figure}

Figure \ref{fig:spectrum_small_diff_rot} shows how good is the $2(2\ell+1)$-Lorentzian model  in reproducing the expected power spectrum for the case of Fig.\ref{fig:noise_1}. The vertical red lines are for the $(2\ell+1)$  peaks with $M=0$,  while the vertical black lines refer to the $(2\ell+1)$ peaks with power  $\langle P\rangle_m $ and frequency shifts $\langle \delta\omega\rangle_m $. The envelope of power of the $2(2\ell+1)$-model compares well with the envelope  obtained by summing over all the $(2\ell+1)^2$ peaks.

Figure \ref{fig:linear_qnl_AR} shows the frequency shifts $\langle \delta\omega\rangle_m $ (black lines) and the shifts of the $M=0$ peaks (red lines) calculated from the  solution of Eq. (\ref{eq:eigen_problem}) (solid lines) and from the approximate solutions (dashed lines), as a function of stellar rotation rate and for an active region with a surface coverage of $4\%~(\alpha=23^\circ)$ and $\beta = 80^\circ$.
The linear approximation successfully returns the frequency shifts of the $2(2\ell+1)$ Lorentzians, even for moderately high rotation rates.

\subsubsection{Limit of small-size active region ($\alpha \lesssim 15^\circ$)}
\label{sec:small_ar}

\begin{figure*}[t]
\includegraphics[width=0.5\columnwidth]{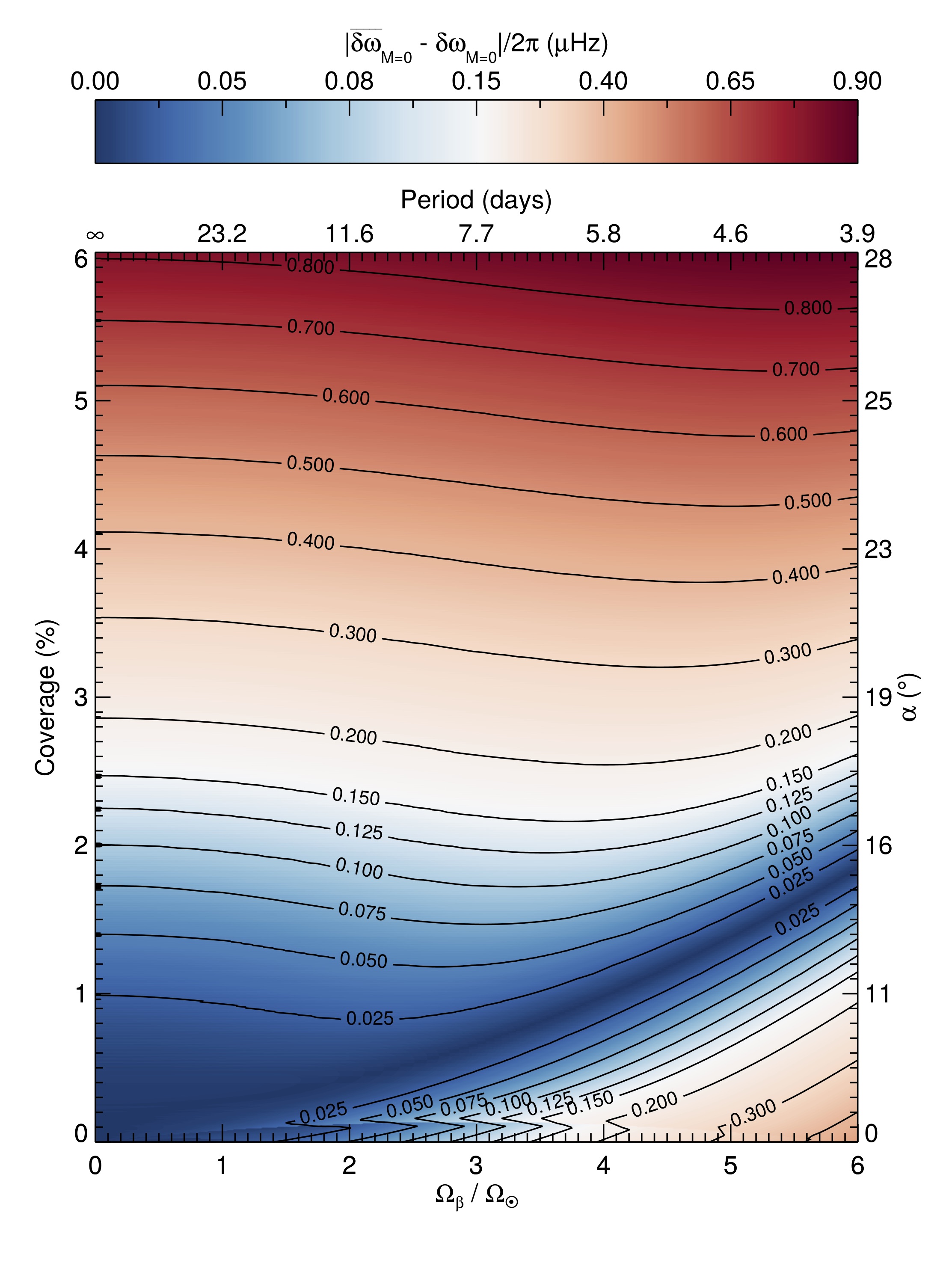}
\includegraphics[width=0.5\columnwidth]{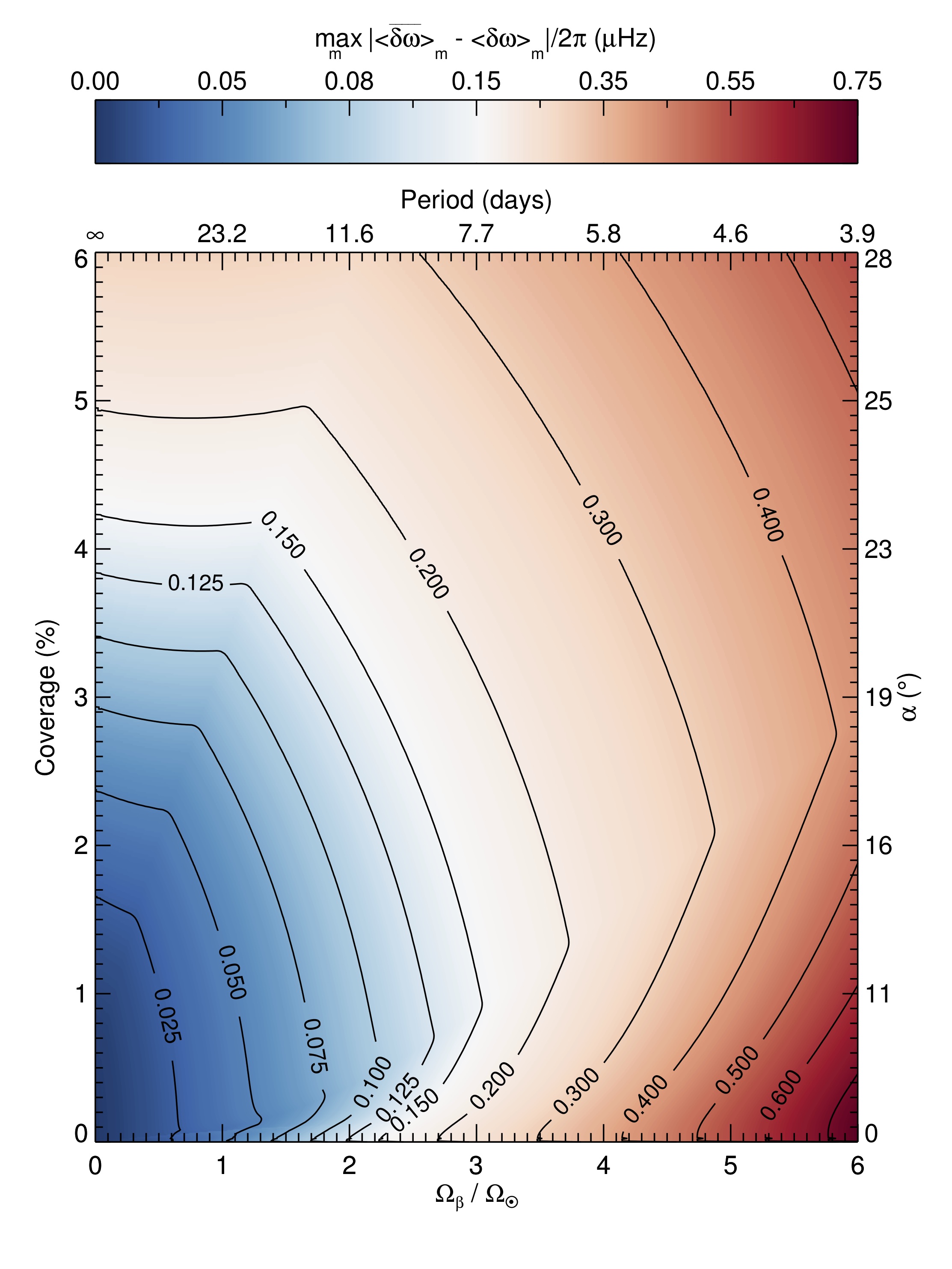}
\caption{
Left: Contour plot of the absolute difference 
$|\overline{\delta\omega}_{M=0} - \delta\omega_{M=0} |$ 
between the approximate shifts $\overline{\delta\omega}_{M=0}$ (see Eq.~\ref{eq:small_ar_shiftsM0}) and the first-order shifts ${\delta\omega}_{M=0}$, as a function of stellar rotation rate and surface coverage of the active region. 
Right: Contour plot of the maximum of the absolute difference
$| \langle \overline{\delta\omega}\rangle_m - \langle \delta\omega\rangle_m | $ over all values of $m$, as a function of stellar rotation rate and surface coverage of the active region.
}
\label{fig:alphaomega_contour}
\end{figure*}

The above formulae for the frequency shifts simplify further when the active region has a small surface area. For small values of $\alpha$, the integral in Eq.~(\ref{eq:glm}) can be approximated,
\begin{align}
\int_{\cos\alpha}^1 \left [P_\ell^{|m''|}(\mu)\right ]^2 \df{\mu}  & 
\underset{\alpha \rightarrow 0}{\sim} 
\left [P_\ell^{|m''|}(0)\right ]^2\frac{\alpha^2}{2} + \OO(\alpha^4)  = \delta_{m''0} \frac{\alpha^2}{2} + \OO(\alpha^4),
\end{align}
such that $G_\ell^{m''}(\alpha)$ becomes
\begin{gather}
\label{eq:Gl0_approx}
G_\ell^{m''}(\alpha) = \delta_{m''0}\frac{\alpha^2}{4} (2\ell+1) + \OO(\alpha^4).
\end{gather}
As shown in Fig.~\ref{fig:glmalpha}, the above approximation is very good for $\alpha \lesssim 15^\circ$ and $\ell \le 2$. 
Up to order $\alpha^2$, the active region induces a shift only in the frequency of the $M=0$ mode
\begin{gather}
\delta\omega_M^{\rm AR} = \delta_{M0}\, \omega_{n\ell}^{(0)} \varepsilon_{n\ell} \frac{\alpha^2}{4} (2\ell+1) + \OO(\alpha^4).
\end{gather}
Perturbed frequency shifts and amplitudes are obtained by following the calculation described in Sect.~\ref{sec:small_diff_rot}, but taking into account the fact that the eigenfrequencies $\delta\omega_M^{\rm AR}$ are now degenerate for $M\neq0$.
The frequencies in the observer's frame can then be approximated by
\begin{gather}
\omega_m^M = \omega_{n\ell}^{(0)} +\overline{\delta\omega}_M +m\Omega_\beta ,
\end{gather}
where 
\begin{align}
\overline{\delta\omega}_M = & \delta_{M0} \, \omega_{n\ell}^{(0)} \varepsilon_{n\ell} \frac{\alpha^2}{4} (2\ell+1) + (1-\delta_{M0})\eta P_2(\cos\beta) Q_{2\ell M}\, \omega_{n\ell}^{(0)}
\,.
\label{eq:small_ar_shiftsM0}
\end{align}
The peak amplitudes in the power spectrum, $P_m^M$, are given by Eq.~(\ref{eq:observed_amplitude}), where $A_m^M$ is replaced by 
\begin{gather}
A_m^M = \sum_{m'=-\ell}^\ell \overline{A}_{m'}^M \; r_{m'm}^{(\ell)}(\beta)
\label{eq:linear_ampl_shifts_smallAR}
  \end{gather}
with
\begin{align}
\overline{A}_{m'}^M = \delta_{Mm'} + (\delta_{M0} - \delta_{m'0})
\frac{\eta\omega_{n\ell}^{(0)}}
{\delta\omega_0^{\rm AR}} \sum_{m''=-\ell}^\ell 
r_{m'm''}^{(\ell)}(\beta) r_{Mm''}^{(\ell)}(\beta) Q_{2\ell m''}.
\end{align}

 Figure~\ref{fig:alphaomega_contour} shows the frequency error in $\mu$Hz introduced by the small-$\alpha$ approximation.
Dark blue shades indicate the regions in parameter space ($\alpha$ and $\Omega_\beta$) where the approximation is very good.
For an active region with a  surface coverage below $2\%$, i.e. $\alpha<15^\circ$, the frequency shift $\overline{\delta\omega}_{M=0}$ is within $\sim 0.1\; \mu$Hz of ${\delta\omega}_{M=0}$, even for fast rotation rates.

\subsection{Correlations in frequency space}
\label{sec:theory_correlations}
Due to the fact that  the active region perturbation is unsteady in the observer's frame, the intensity fluctuations (Eq.~\ref{eq:Iobs_final}) are not statistically independent in frequency space. 
Given two frequencies $\omega$ and $\omega'$, the intensity covariance  is 
\begin{align}
 \text{Cov} & \left [ {I_\text{obs}}(\omega), \, I_\text{obs}(\omega')\right ] \nonumber\\
&
= E\left [ {I^*_\text{obs}}(\omega) I_\text{obs}(\omega')\right ] \nonumber \\ 
& =\sum_{M,m} \sum_{M',m'}  B_m^M  B_{m'}^{M'} L^{1/2}_M(\omega-m\Omega_\beta)  L^{1/2}_{M'}(\omega'-m'\Omega_\beta)
E[\mcN^*_M(\omega- m\Omega_\beta) \mcN_{M'}(\omega'- m'\Omega_\beta)]
\nonumber\\
& =\sum_{M=-\ell}^{\ell} \sum_{m=-\ell}^{\ell} \sum_{m'=-\ell}^{\ell} B_m^M B_{m'}^M L_M(\omega-m\Omega_\beta)\delta_{m',m+(\omega'-\omega)/\Omega_\beta}   \,  .
\label{eq:I_correlation}
\end{align}
To simplify the analysis we assumed that $k=(\omega'-\omega)/\Omega_\beta$ is an integer (this is not a weakness of the theory though). The above expression vanishes unless   $| k | \le 2 \ell $.  The quantities $I_\text{obs} (\omega)$ and $I_\text{obs} (\omega')$ are correlated for frequency separations $\Delta\omega = \omega'-\omega=(m'-m)\Omega_\beta$.

The power spectrum  is also correlated for  
$\Delta \omega =(m-m')\Omega_\beta$. Using the formulae given in appendix C of \citet{fournier2014}, we find
\begin{align}
 \text{Cov}\left [ P(\omega),P(\omega') \right ] 
 &= 
 \text{Cov}\left [ I^*_\text{obs}(\omega)I_\text{obs}(\omega) , \, I^*_\text{obs}(\omega')I_\text{obs}(\omega') \right ]  \nonumber \\
 & = E \left [ I^*_\text{obs}(\omega)I_\text{obs}(\omega')  \right ]  E \left [ I_\text{obs}(\omega)I^*_\text{obs}(\omega')  \right ]  \nonumber \\
&= \left | \text{Cov} \left [ I_\text{obs}(\omega), \, I_\text{obs}(\omega') \right ] \right |^2  ,
\label{eq:PS_correlation}
\end{align}
where the intensity covariance is given above.
We note that the values of the intensity and power spectrum covariances  (Eqs.~\ref{eq:I_correlation} and~\ref{eq:PS_correlation}) depend on the parameters of the model. However the existence of a  correlation is a general feature, which arises from the fact that the active region rotates with the star.   
Remarkably, the expectation value  of the power spectrum $\mc{P}(\omega)$ (see Eq.~\ref{eq:powerspectrum})  is as if all the terms in Eq.~(\ref{eq:Iobs_final}) were statistically independent.

\section{Discussion}
\subsection{Nonlinear frequency shifts and amplitudes from numerical simulations}
\label{cha:rotation:sec:simulation}

In order to compare with the perturbation theory, we study the nonlinear regime by means of numerical simulations.
We use the GLASS wave propagation code, with the same numerical setup employed by  \citet{Papini2015}.

\subsubsection{Rotation in post-processing}
Running different numerical simulations for different values of $\beta$ and different perturbation amplitudes is computationally expensive. 
Instead we performed simulations for a 3D polar perturbation to the sound speed and for a star with no rotation, that is equivalent to solve the numerical problem in the reference frame $\wmcR_\beta$.
We introduced the effect of rotation later in processing the output.
This approach has the advantage that, for a given amplitude of the AR perturbation, we only need to run one simulation in order to calculate the power spectrum for any given value of $\beta$ and rotation period.
However we can only reproduce solid body rotation, and it is not possible to include the effects of the centrifugal distortion and of the Coriolis force,
therefore in each $n\ell$-multiplet we expect to find only $(2\ell+1)(\ell+1)$ peaks. 
Nonetheless the results are useful for exploring the nonlinear regime.
We note that, as a consequence of neglecting these rotational effects, $M$ can be identified with the azimuthal degree of the spherical harmonics $Y_\ell^M (\widehat\theta_\beta,\widehat\phi_\beta) $ in the frame $\wmcR_\beta$ (see Sect.~\ref{sec:corot}).
\subsubsection{Sound-speed perturbation}
As was done earlier, we approximate the perturbation introduced by the starspot by a local increase in sound speed.
We consider separable perturbations in the square sound speed of the form
\begin{equation}
\label{eq:deltacs2spot}
 {\Delta c^2(r,\widehat\theta_\beta)} = \mathrm{\epsilon} \, c_0^2(r)  \, f(r) \, g(\widehat\theta_\beta), 
 \end{equation}
 where $\epsilon>0$ is a positive amplitude, $f$ is a radial profile, and $g$ a latitudinal profile.
Explicitly, $g = 1/2 + \cos(\kappa \theta)/2$ is a raised cosine for $0\le \kappa\theta<\pi$ and is zero otherwise, where $\pi/(2\kappa)= 0.65$~rad~$=37.5^\circ$. The function $f=\exp(-|r-r_c|^2/2\sigma^2) [1/2 + \cos(|r-r_c|/\sigma)/2]$  is a Gaussian function centered at radius $r_c=0.9985~R_\odot$ with dispersion $\sigma=0.004~R_\odot$, multiplied by a raised cosine. 
This functional form is the same as the one described by \citet{Papini2015}.
The perturbation is thus placed along the polar axis at a depth of $4$~Mm, with a surface coverage of $12\%$.  

\subsection{Synthetic power spectrum}

\label{cha:rotation:sec:simulationpowerspectrum}
As in Sect.~\ref{sec:linear_PS}, we assume that the intensity fluctuations are proportional to the Eulerian pressure perturbation measured at $r_0=R+200~\mathrm{km}$ above the surface \citep[see ][]{Papini2015}. 
To calculate the expectation value of the observed power spectrum, we performed a first set of simulations for which all the modes were excited with the same phase at the initial time.

The approximation that the changes in the eigenfunction of a mode $M$ are limited to the same angular degree $\ell$ (Eq. \ref{eq:xi_pert}) does not hold in the nonlinear case: the horizontal shape of an eigenfunction is a combination of spherical harmonics of different $\ell$ values. 
However, since the perturbation is axisymmetric (see Eq.~\ref{eq:deltacs2spot}), 
the eigenfunction for a mode $M$ is a combination of spherical harmonics  $Y_\ell^M(\widehat\theta_\beta,\widehat\phi_\beta)$ of different angular degree $\ell$ and same $M$.
Therefore the intensity fluctuations  $I_{M}(\widehat\theta_\beta,\widehat\phi_\beta,t)$ due to all the modes with the same $M$ take the form
\begin{gather}
\label{eq:I_Mglass}
 I_{M}(\widehat\theta_\beta,\widehat\phi_\beta,t) \propto \sum_{\ell=|M|}^{\ell_\text{max}}\Re \left \{ 
 {p}_{\ell M}(r_0,t)  
 Y_\ell^M(\widehat\theta_\beta,\widehat\phi_\beta) \right \},
\end{gather}
where ${p}_{\ell M}(r_0,t)$ are the coefficients of the spherical harmonic decomposition of the pressure wavefield $p(r_0,\widehat\theta_\beta,\widehat\phi_\beta,t)$  returned by the GLASS code  in the frame $\widehat\mcR_\beta$. The index $\ell_\mathrm{max} $ is set by the spectral resolution of the spherical harmonic transform.

To obtain the full-disk integrated intensity in the observer's frame 
we express each $Y_\ell^M(\widehat\theta_\beta,\widehat\phi_\beta)$ in terms of the spherical harmonics in the frame $\mcR$ 
\begin{gather}
Y_\ell^{M}(\widehat\theta_\beta,\widehat\phi_\beta) = 
\sum_{m=-\ell}^{\ell} 
r_{mM}^{(\ell)}(\beta) e^{-\mathrm{i} m\Omega_\beta t} 
\sum_{m'=-\ell}^{\ell}
Y_\ell^{m'}(\theta,\phi) \; r_{m' m}^{(\ell)}(-i) 
  .
\end{gather}
by means of two consecutive rotations of the Euler angles $(0,-i,\Omega_\beta t)$ and $(0,\beta,0)$. Combining this equation with  Eq.~(\ref{eq:I_Mglass}) and integrating over the visible disk, we obtain the full-disk integrated intensity of each $mM$-component:
\begin{gather}
\label{eq:integrated_intensity_sim}
I_{mM}(t)= 
\sum_{\ell=\max\{|m|,|M|\}}^{\ell_{\max}} V_\ell\; \rl_{m0}(i) \, \rl_{mM}(\beta) \, 
\Re \left \{{p}_{\ell M}(r_0,t)  e^{-im\Omega_\beta t} \right \}.
\end{gather}
We then perform a Fourier transform to calculate the intensity $I_{mM}(\omega)$ in the frequency domain. Finally, we derive the expression for the expectation value of the  power spectrum
\begin{gather}
\label{eq:powerspectrum_sim}
\mc{P}(\omega)=  \sum_{m=-\ell_{\max}}^{\ell_{\max}}\sum_{M=-\ell_{\max}}^{\ell_{\max}} 
\left |I_{mM}(\omega)\right |^2,
\end{gather}
that is analogous to Eq. (\ref{eq:powerspectrum}), but for the entire wavefield.

\begin{figure*}
 \includegraphics[width=0.5\columnwidth]{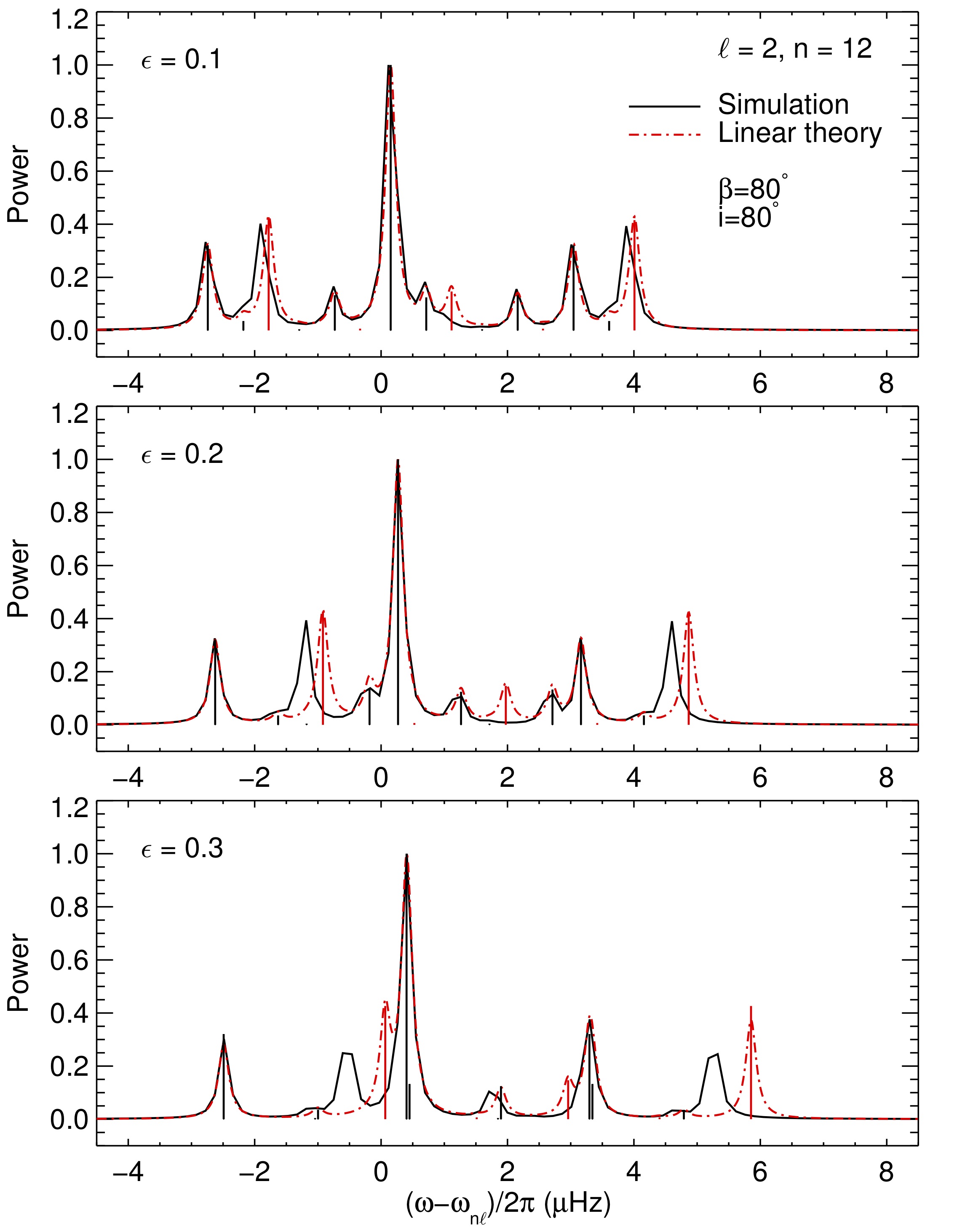}
 \includegraphics[width=0.5\columnwidth]{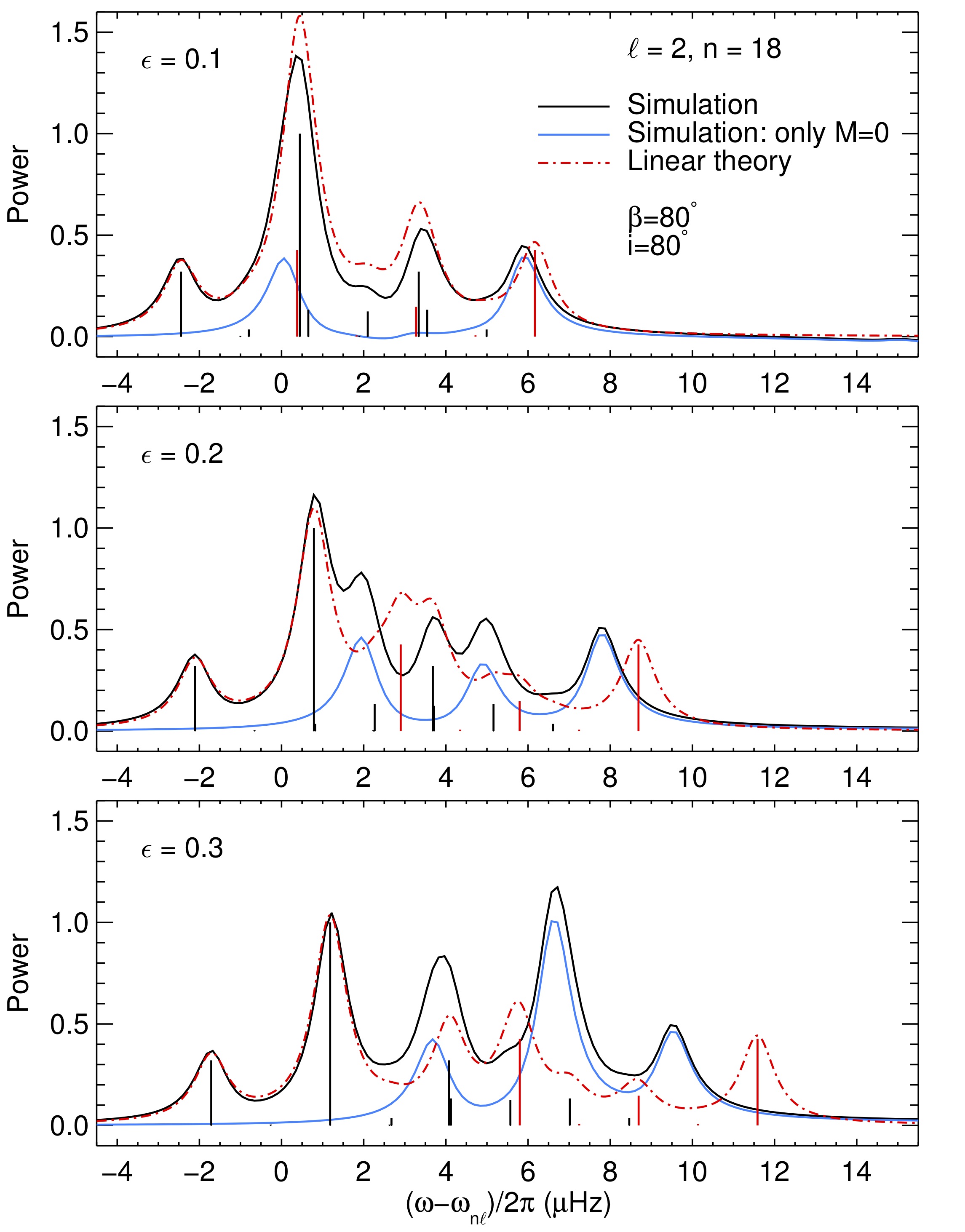}
 \caption{Numerical simulations of oscillation power spectra for two quadrupole multiplets with $n=12$ (left panels) and $n=18$ (right panels), for an inclination angle $i=80^\circ$ and a 
stellar  rotation period of $8~\text{days}$ (solid body rotation).
The active region has a colatitude  $\beta=80^\circ$ and different perturbation amplitudes of $\epsilon=0.1$ (top), $0.2$ (middle), and $0.3$ (bottom), with $r_c=0.9985~R_\odot$, $\sigma = 0.004~R_\odot$, and $\kappa=2.4$ (see Eq. \ref{eq:deltacs2spot}).
 The black curves show the power spectra computed with  the GLASS code, the red dot-dashed curves indicate the power spectra computed using  linear perturbation theory. Vertical lines show the  peaks from linear theory, in red for $M=0$ and in black $M\neq0$.
 The blue curves in the right panels display the contribution of the $M=0$ peaks to the simulated power spectrum.
 The $n=12$ peaks  have a FWHM of $\Gamma/2\pi\simeq0.2~\mu\Hz$, while for $n=18$ the peaks have   $\Gamma/2\pi\simeq1~\mu\Hz$.
 }
 \label{fig:l2spectrum_sim}
\end{figure*}

For the nonlinear study we chose three perturbation amplitudes   $\mathrm{\epsilon}=0.1$, $0.2$, and $0.3$ which, for the multiplet $(\ell,n)=(2,18)$, correspond to $\epsilon_{n\ell}\simeq 0.005$, $0.010$, and $0.015$, \ie, roughly twice to six times the value used in the linear analysis.
The simulations run for $80~\text{days}$ (stellar time), in order to reach an accuracy of $\sim 0.14~\mu\Hz$ in the frequency domain.
The wavefield computed by GLASS includes some numerical damping that increases with frequency with an exponential dependence.
We took advantage of this damping and  selected two $\ell=2$ multiplets: one with $n=18$ and a FWHM comparable to the solar value, the other with $n=12$ and a FWHM small enough to resolve all the $mM$ peaks in the multiplet.  
Model~S is convectively stabilized \citep{Papini2014}, which implies that the unperturbed frequencies of the quadrupole modes are $1970.50$~$\mu$Hz for $n=12$ and $2783.62$~$\mu$Hz for $n=18$.
Figure \ref{fig:l2spectrum_sim} shows the observed power spectra of the two selected multiplets in the case   $\beta=80^\circ$ and $i=80^\circ$, normalized with respect to the highest $M\neq0$ peak.
In the case $\epsilon=0.1$, the simulated power spectrum (black curve) of the $(\ell,n)=(2,12)$ multiplet (top left panel) is well reproduced by the power spectrum computed with linear theory (red dot-dashed curves), except for the peaks corresponding to $M=0$ (vertical red curves show the $M=0$ peaks from linear theory), which are less shifted in frequency and have smaller amplitudes than predicted. 
For the $(\ell,n)=(2,18)$ multiplet (top right panel) the nonlinear effects are less visible, due to the overlapping  Lorentzian profiles. A blue curve, displaying the contribution to the power spectrum of the $M=0$ peaks, shows that also for this multiplet the $M=0$ component of the power spectrum deviates from the linear behaviour, both in frequency and amplitude. 
 This plot also shows an example of the combined action of mode mixing and mode visibility, which almost suppresses the $m=0$, $M=0$ peak located at a frequency shift of $\sim 3.1~\mu\Hz$.

With increasing $\epsilon$ (middle and bottom panels) the interaction of the wavefield  with the active region becomes strongly nonlinear, and for a perturbation with $\epsilon=0.3$ results in a massive distortion of the power spectrum with respect to that  one predicted by linear theory. 
Here two different behaviors are evident: in the $(\ell,n)=(2,12)$ multiplet the $m=0, M=0$  peaks are almost suppressed, while  
the peak with $m=0, M=0$ in the $(\ell,n)=(2,18)$ multiplet, that was suppressed in the case $\epsilon=0.1$, increases in amplitude as $\epsilon$ increases (middle and bottom right panels).
Moreover the peaks with $M=1$ start to deviate from the linear prediction.
This is in agreement with what found in the non-rotating case by \citet{Papini2015}, who also showed that second-order perturbations would correct most of the differences.

\subsection{Correlations in synthetic power spectra}

\begin{figure}
  \includegraphics[width=\columnwidth]{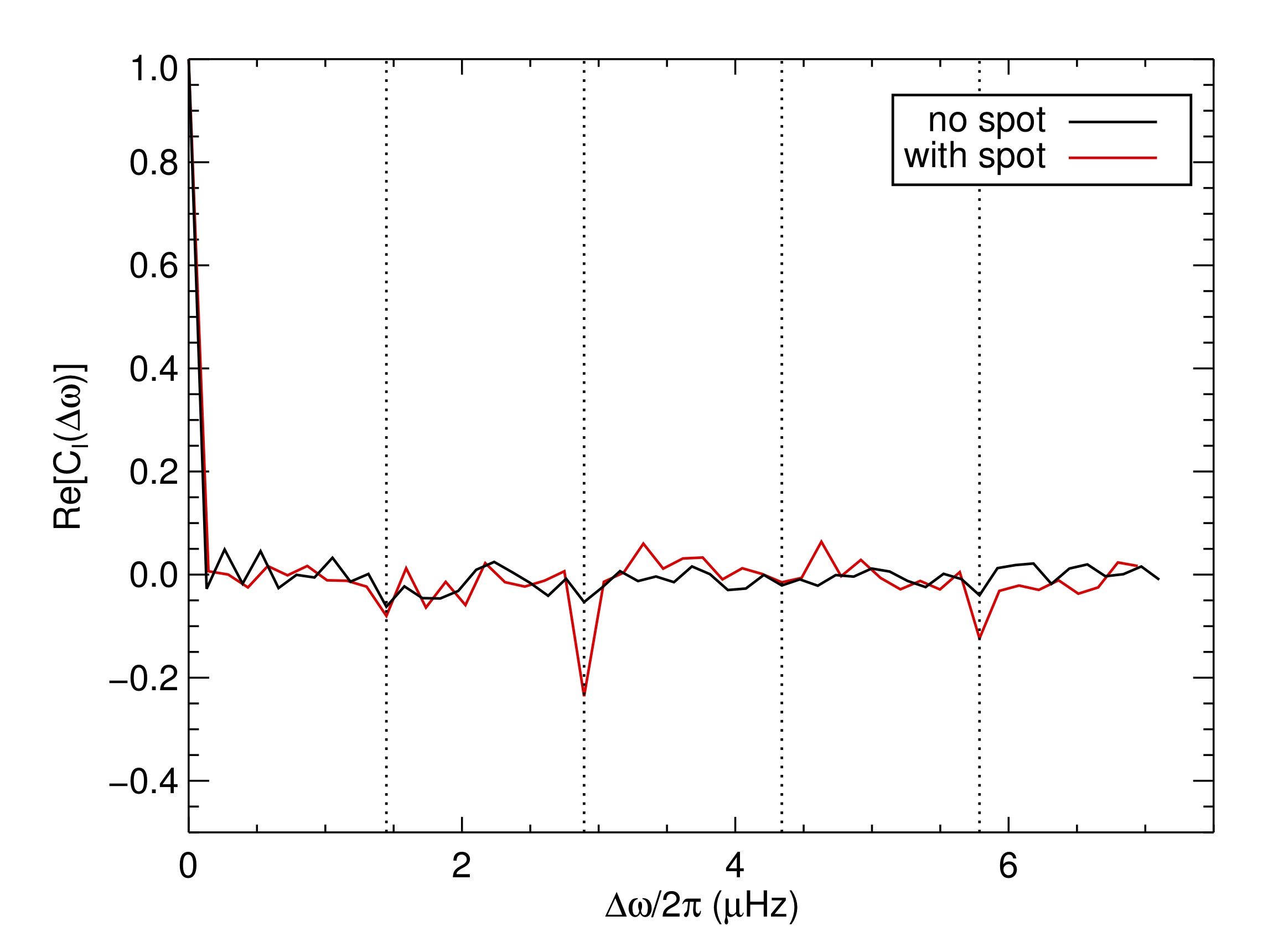}
 \caption{ Autocorrelation of the intensity spectrum as defined by  Eq. (\ref{eq:autocorrelation_PS_I}) for a starspot with $\epsilon=0.1$ (red curve).
  A stellar rotation period $2\pi/\Omega_\beta = 8$~days, an inclination angle $i=80^\circ$, and a starspot colatitude $\beta=80^\circ$ were chosen for the post processing. Vertical dotted lines denote frequency separations $\Delta\omega=j\Omega_\beta$ where $j=1,2,3, 4$.  For comparison, the black curve is the case with no starspot.
  }
  \label{fig:correlation_glass}
 \end{figure}
 
In Sect.~\ref{sec:theory_correlations} we showed that in presence of an active region rotating with the star, the power spectrum of a multiplet is correlated at frequency separations that are multiple of the rotational frequency of the active region. However, for small perturbations this correlation is too weak to be observed. Here we ask whether such a correlation can be measured in presence of a perturbation of moderate amplitude. 
For that purpose, we ran a second set of simulations, in which the acoustic waves were excited by applying a random forcing function at  $150~\text{km}$ below the surface at each time step, as described by \citet{Hanasoge2007}.
The duration of the simulation is 80 days (stellar time).
In the frequency domain, the observed intensity is 
\begin{equation}
I_\mathrm{obs} (\omega) = \sum_{m=-\ell_{\max}}^{\ell_{\max}}\sum_{M=-\ell_{\max}}^{\ell_{\max}} I_{mM}(\omega) ,
\end{equation}
where $I_{mM}(\omega)$ is obtained by Fourier transformation of  Eq.~(\ref{eq:integrated_intensity_sim}) using the numerical realizations of $p_{\ell M}(r_0, t)$.

The autocorrelation of  the intensity spectrum is
\begin{gather}
 C_I(\Delta\omega) = 
 \frac{\sum_{n=15}^{25}  \int_{[\omega_{n\ell}]}  
 I_\text{obs}^*(\omega) I_\text{obs}(\omega+\Delta\omega)\df{\omega} }
 {\sum_{n=15}^{25} \int_{[\omega_{n\ell}]}  
 |I_\text{obs}(\omega)|^2 \df{\omega} } ,
 \label{eq:autocorrelation_PS_I} 
\end{gather}
where $[\omega_{n\ell}]$ denotes an appropriate frequency interval of size $\sim20~\mu$Hz containing the multiplet $n\ell$ but excluding the nearby $l=0$ mode. 
The average is performed over all $\ell=2$ multiplets with $n$ ranging from 15 to 25.

Figure \ref{fig:correlation_glass} shows the real part of $C_I$  for the special case of an inclination angle  $i=80^\circ$ and a starspot at colatitude  $\beta=80^\circ$ with $\epsilon=0.1$.
A correlation is clearly visible at frequency separations $\Delta\omega=2\Omega_\beta$ and  $4\Omega_\beta$.
This suggests that the frequency-domain  autocorrelation function could be used as a diagnostic tool to identify 
unsteady perturbations in the time series of stellar oscillation. 
We note that the imaginary part of $C_I$ contains no visible signal above the noise level.

\subsection{Towards a physical model for mode interaction with an active region}
\label{SectPhys}

{ 
In this paper we replaced the  active region  by a localized  increase in sound-speed near the stellar surface.
We focused on the geometrical aspects of the problem rather than on the physics.
One may ask, however,  what would be the difference in the obtained results if we had instead considered a realistic model for the magnetic active region.
Although we will not answer this question here, it is worth listing some of the steps involved.

A typical solar active region consists of a pair of sunspots surrounded by plage with strong vertical field. 
Local helioseismology of the visible disk and the far side indicates that p modes are strongly scattered by both sunspots and extended plage \citep[see, e.g.,][and references therein]{Gizon2009}.
Some studies of the interaction of high-degree p~modes with magnetostatic sunspots  \citep[e.g.,][]{Moradi2010,Cameron2011} have been carried out using MHD wave propagation codes \citep{Cameron2008,Felipe2016}. 
Other studies are based on  numerical magneto-convective simulations \citep[e.g.,][]{Rempel2009,DeGrave2014}. 
The main conclusion of these simulations is that the interaction takes place in the top few hundred kilometres below the surface, where the direct effects of the magnetic field and the indirect effects due to changes in thermodynamic structure with respect to the reference atmosphere (e.g. the Wilson depression) are large.
Wave simulations indicate that the outgoing p modes are phase shifted with respect to the incoming p modes in such a way that the effective wave speed is increased, as observed. 
The physical interaction involves the conversion of p modes  into fast and slow magnetoacoustic modes in the sunspot  \citet[e.g.][]{Cameron2008,Khomenko2006}. A fraction of the incoming p-mode  energy is tunnelled downward in the form of slow MHD waves, leading to absorption  \citep{Braun1995,ZhaoH2016}. See also, e.g.,  \citet{Saio2004} and  \citet{Cunha2006} for mode conversion calculations in roAp stars.

Using 2D ray tracing, \citet{Liang2013} showed that high-degree helioseismic waveforms can be reproduced by increasing the effective wave speed by 10\%  in the sunspot.
 This provides some justification for the values that we have used in the present paper, although the extension to low-degree p modes has not been studied.
Clearly, much additional work will be needed to determine the correct active-region perturbation amplitude from first principles. Until then, a simple calibration can be obtained from the observational study by \citet{Santos2016} who estimated empirically the contribution of sunspots to the low-degree p-mode frequency shifts associated with the solar cycle. By  combining our Eq. (\ref{eq:omat_mag_epsilon}) with Eq. (3) from \citet{Santos2016}, we find  $\varepsilon_{n\ell}=-\Delta\delta_\mathrm{ch}/I_\ell$, where $\Delta\delta_\mathrm{ch}$ is the integral phase difference introduced in the mode eigenfunction by a sunspot and $I_\ell$ is related to mode inertia. Using  the value $\Delta\delta_\mathrm{ch}=-0.44$  estimated by \citet{Santos2016}, we have  $\varepsilon_{n\ell} \sim 0.05$ for a quadrupole mode, \ie\  a larger value than proposed by \citet{Liang2013} and used in the present paper. This may suggest that for large active regions the spectra calculations may have to be carried out in the nonlinear regime. However, only realistic numerical modelling would help settle this question. The full problem would also have to include multiple scattering by collections of flux tubes in plage  \citep[see, e.g.,][]{Hanson2015}. 
}

\section{Conclusions}

In this paper we investigated the changes in global acoustic oscillations caused by a localised sound-speed perturbation on a rotating star mimicking  a large active region, using both linear perturbation theory and 3D numerical simulations. In an inertial frame, the active region perturbation is unsteady.

We find that the power spectra of low-degree modes have a complex structure. The combined effects of the active region and differential rotation cause each $n\ell$-multiplet to appear as $(2\ell+1)^2$ peaks, each with a different amplitude. 
Most of the peaks are clustered near the classical rotationally-split frequencies, and only $2\ell+1$ peaks (the $M=0$ peaks, which correspond to the axisymmetric mode in the reference frame of the AR) are shifted to higher frequencies. This leads to an apparent asymmetry in the line profiles. However, due to the finite lifetime of acoustic oscillations, most of the peaks cannot be resolved.
{ For solar-type stars, the results are not very sensitive to the choice of latitudinal differential rotation profile.}

The structure of the power spectra  is sensitive to the latitudinal position of the active region and to the inclination angle $i$ of the stellar rotation axis. The latter plays a major role in determining the relative visibility of the individual peaks.
We find that the envelope of the power spectrum becomes more complex as the latitude of the active region decreases. 
In practice, it would be very difficult to perform a fit of the $(2\ell+1)^2$ peaks in a multiplet, due to peak blending and noise. However, by neglecting differential rotation it is possible to derive a simplified formula that approximates the power spectrum of a multiplet to a sum of only $2(2\ell+1)$ Lorentzian profiles. For small-area  active regions, the formula further simplifies and directly links the frequencies of the peaks in the power spectrum  to the active region parameters. Such formula may find applications in the analysis of real asteroseismic observations.

Numerical simulations were performed to explore the nonlinear regime of the perturbation. We find that the $M=0$ peaks deviate from the linear behavior for active-region perturbation amplitudes $\varepsilon_{n\ell} \gtrsim 0.005$. Depending on  each particular case, the amplitude of these peaks is either reduced or enhanced compared to first-order linear theory, due to mixing  with modes with other values of $\ell$ and $m$ \citep{Papini2015}.

We found that there are correlations in the power spectrum at frequency separations that are multiples of the active-region rotation rate. In the linear regime the correlation signal is too weak to be observed. However the numerical simulations show that for active-region perturbations of moderate amplitude, such a correlation might be detectable, provided that  the frequency intervals are carefully selected  to increase the signal to noise ratio.

We note that perturbation theory can easily be extended to compute the effect of multiple active regions provided that latitudinal differential rotation is small. The  treatment of  several active regions rotating at different rotation rates  would require a different setup, since there is no frame in which these perturbations are steady. A numerical approach would be preferable in such a case.

{ The work presented in this paper uses simplified physics, but  it should provide useful guidance to identify the seismic signature of a large active region in the power spectrum of stellar oscillations. 
Given that the values of $\epsilon_{n\ell}$ are uncertain, we believe that it is worth searching  for low-degree multiplets consisting of $2(2\ell+1)$ components in available asteroseismic observations.}
Ideal targets are stars that are known to have high-quality oscillation power spectra \citep[high SNR, narrow line profiles, clear rotational splitting, see e.g.][]{Nielsen2014} 
and  show evidence for long-lived starspots \citep[e.g.][]{Nielsen2013, Nielsen2018}. 
The catalogue of potential targets, currently limited to CoRoT and {\it Kepler},  will increase fast with  TESS \citep{TESS2015} and  PLATO \citep{plato2014}.

\section*{Conflict of Interest Statement}

The authors declare that the research was conducted in the absence of any commercial or financial relationships that could be construed as a potential conflict of interest.

\section*{Author Contributions}

L.G. provided the basic theory and E.P. the numerical applications. Both authors contributed to the analysis of the results and to the writing of the article.

\section*{Acknowledgments}

This paper is a contribution to PLATO PSM activity `Seismic diagnostics of stellar activity'. We acknowledge financial support from the German Aerospace Center DLR and from the Max Planck Society.
We thank Shravan Hanasoge for making the GLASS code available. 
Computational resources were provided by the German Data Center for SDO.
This work appeared in part in the PhD thesis of \citet{Papini-thesis}. 

\bibliographystyle{frontiersinSCNS_ENG_HUMS} 
\bibliography{biblio}

\end{document}